\documentclass[12pt,preprint]{aastex}

\begin{document}

\title{The Globular Cluster Populations of Giant Galaxies: Mosaic Imaging of Five Moderate-Luminosity Early-Type Galaxies} 

\author{Jonathan R. Hargis and Katherine L. Rhode} 
\affil{Indiana University, 727 East 3rd Street,
Swain West 319, Bloomington, IN 47405,USA}
\email{jhargis@astro.indiana.edu,rhode@astro.indiana.edu}

\shorttitle{Globular Cluster Systems of Moderate-Luminosity Giant Galaxies}
\shortauthors{Hargis and Rhode}

\received{}
\accepted{}


\begin{abstract}

This paper presents results from wide-field imaging of the globular
cluster (GC) systems of five intermediate-luminosity ($M_V ~\sim -21$
to $-22$) early-type galaxies.  The aim is to accurately quantify the
global properties of the GC systems by measuring them out to large
radius. We obtained $BVR$ imaging of four lenticular galaxies
(NGC~5866, NGC~4762, NGC~4754, NGC~3384) and one elliptical galaxy
(NGC~5813) using the KPNO 4m telescope and MOSAIC imager and traced
the GC population to projected galactocentric radii ranging from $\sim
20$ kpc to $120$ kpc.  We combine our imaging with {\it Hubble Space
Telescope} data to measure the GC surface density close to the galaxy
center.  We calculate the total number of GCs ($N_{\rm GC}$) from the
integrated radial profile and find $N_{\rm GC}=340 \pm 80$ for
NGC~5866, $N_{\rm GC}=2900 \pm 400$ for NGC~5813, $N_{\rm GC}=270 \pm
30$ for NGC~4762, $N_{\rm GC}=115 \pm 15$ for NGC~4754, and $N_{\rm
GC}=120 \pm 30$ for NGC~3384.  The measured GC specific frequencies
are $S_N$ between 0.6 and 3.6 and $T$ in the range 0.9 to 4.2. These
values are consistent with the mean specific frequencies for the
galaxies' morphological types found by our survey and other published
data.  Three galaxies (NGC~5866, NGC~5813, NGC~4762) had sufficient
numbers of GC candidates to investigate color bimodality and color
gradients in the GC systems.  NGC~5813 shows strong evidence
($>3\sigma$) for bimodality and a $B-R$ color gradient resulting from
a more centrally concentrated red (metal-rich) GC subpopulation.  We
find no evidence for statistically significant color gradients in the
other two galaxies.
\end{abstract}

\section{Introduction}\label{sec_intro}

Common to all types of galaxies -- from dwarfs to giant cD galaxies -- is a population of globular star clusters (GCs). Studies of the Milky Way GC population have shown GCs to be among the oldest known stellar populations ($\sim 11-13$ Gyr; \citealt{kr03,fo10} and references therein), and as such these objects provide an observable record of the chemical and dynamical history of the Galaxy \citep{se78,zi93,as98}.  As luminous, compact objects (average $M_V \sim -7$, median half light radii $\sim 3$ pc; \citealt{as98}), GCs are easily observable out to large distances ($\sim 20-30$ Mpc) and are therefore important probes of galaxy formation and evolution outside the Local Group (see \citealt{br06} and references therein).  

In recent years, the GC systems of giant galaxies have been investigated using high-resolution, hierarchical galaxy formation simulations in a cosmological context \citep{be02,kr05,mo06,bo09,gr10,mu10}. Critical to testing such simulations are accurate measurements of the global properties of galaxy GC systems: total numbers of clusters, specific frequencies, spatial distributions, and color (metallicity) distributions.  We have been carrying out a wide-field imaging survey of GC systems of giant spiral, elliptical, and lenticular galaxies outside the Local Group ($d \sim 15-20$ Mpc).  The survey design is presented in Rhode \& Zepf (2001; hereafter RZ01) and previous results are given in Rhode \& Zepf (2001, 2003, 2004; hereafter RZ01, RZ03, RZ04), Rhode et al. (2005, 2007, 2010; hereafter R05, R07, R10), Hargis et al. (2011; hereafter H11), and Young, Dowell, \& Rhode (2012; hereafter Y12).  The primary aim of our survey has been to characterize the global properties of GC systems of giant galaxies.  Such measurements -- made over a range of host galaxy luminosity, morphology, and environment -- provide fundamental observational constraints to galaxy formation models.

Studies of the GC systems of moderate-luminosity galaxies in particular ($M_V \sim -21$ to $-22$) often lack sufficient spatial coverage to provide directly measured ensemble GC system properties.  Giant galaxy GC systems often extend to 50-100 kpc from the galaxy center ($10\arcmin$ to $20\arcmin$ at 15 Mpc), and therefore wide-field observations are necessary to directly measure these properties and trace GC systems over their full radial extent.  A significant number of galaxy GC systems, covering a wide range in host galaxy luminosity, have been studied with the  {\it Hubble Space Telescope} ({\it HST}; C{\^o}t{\'e} et al. 2004, Jord{\'a}n et al. 2007, Kundu \& Whitmore 2001a,b). 
However, the limited spatial coverage of the widest-field {\it HST} instrumentation provides radial coverage of only $\sim 11$ kpc for galaxies at Virgo Cluster distances ($\sim 3\farcm4\times3\farcm4$ for the Advanced Camera for Surveys Wide Field Camera; ACS/WFC).  Therefore, only lower luminosity galaxies ($M_V \sim -21$ or fainter) will have full radial coverage of their GC systems in studies with {\it HST}.  In contrast, wide-field imaging studies have typically focused on the most luminous ($M_V \sim -22$ or brighter) galaxies, e.g., NGC~4472, NGC~4406 (Rhode \& Zepf 2001, 2004); NGC~1399 (Dirsch et al. 2003); NGC~1407 (Forbes et al. 2011); NGC~5128 (Harris et al. 2004); M87 (Tamura et al. 2006).  Consequently, studies of the GC systems of intermediate-luminosity galaxies with adequate spatial coverage remain rare in the literature.
 
This paper presents the results of wide-field imaging study of the GC systems of five intermediate-luminosity ($M_V ~\sim -21$ to $-22$) early-type galaxies: one elliptical and four lenticular (S0) galaxies. The basic properties of the target galaxies are given in Table~\ref{tbl_targets}. We obtained $BVR$ imaging using the Kitt Peak National Observatory (KPNO) Mayall 4m telescope and Mosaic camera and, where possible, complemented this data with published and archival data from the Hubble Space Telescope ({\it HST}). The galaxies have been chosen both to increase the number of moderate-luminosity galaxies with well measured global GC system properties and to fill in existing gaps in our survey.  In particular, this study doubles the number of S0 galaxies in the survey, resulting in comparable numbers of spiral, elliptical, and S0s in the  sample.  Hargis \& Rhode (2013, in preparation) will present a multivariate analysis that will include the five galaxies in this study as well as the $\sim 25$ other galaxies analyzed for the survey to date.

While each target galaxy has had a previous GC system study, this work is the first {\it wide-field} CCD study of NGC~5866, NGC~5813, NGC~4762, NGC~4754, and NGC~3384.  Given the sparseness of this galaxy's group environment, the lenticular NGC~5866 ($M_V = -21.1$) has been characterized as a field galaxy (see Li et al. 2009 and references therein).  The GC system was previously studied by Cantiello, Blakeslee, \& Raimondo (2007; hereafter CBR07) who used archival {\it HST} ACS data to study the color distribution of the GC system. The elliptical galaxy NGC~5813 ($M_V -22.3$) is the second brightest member of the NGC~5846 galaxy group \citep{ma05}.  Previous studies of the total number of GCs in NGC~5813 were done by \citet{ha81} and \citet{ho95}.  \citet{ha81} used single-filter photographic plate data to estimate the specific frequency and total number of GCs.  \citet{ho95} used $BVR$, small-format CCD imaging (three $4\farcm3 \times2\farcm7$ pointings) of NGC~5813 to estimate the total number of GCs and study the color distribution of the galaxy's GC system. The edge-on lenticular galaxy NGC~4762 ($M_V = -21.2$) is located on the eastern edge of the Virgo Cluster and was noted by \citet{sa61} as having ``the flattest form of any galaxy known".   The lenticular galaxy NGC~4754 ($M_V = -20.7$) forms a wide but non-interacting pair with NGC~4762 \citep{RC1}.  Both galaxies were included in the ACS Virgo Cluster Survey (C{\^o}t{\'e} et al. 2004; hereafter ACSVCS).  The total numbers of GCs and color distributions of NGC~4762 and NGC~4754 were studied by \citet{pe06,pe06b,pe08}.  A member of the Leo I group of galaxies, the lenticular galaxy NGC~3384 ($M_V = -20.5$) forms a close pair (in projection) with NGC~3379 but the two galaxies are non-interacting \citep{RC2}.  \citet{ha81} estimated the specific frequency and total number of GCs in their single-filter photographic plate study of the NGC~3379 field.

This paper is organized as follows.  Section~\ref{sec_observations} describes our observations and data reduction procedures.  Section~\ref{sec_detection} presents our methods and results for the detection of the GC systems.  Section~\ref{sec_analysis} describes our analysis of the GC systems. Section~\ref{sec_global} presents our results on the global properties of the GC systems.  Section~\ref{sec_conclusions} gives a summary of our results.

\section{Observations and Data Reduction}\label{sec_observations}

\begin{figure}
\plotone{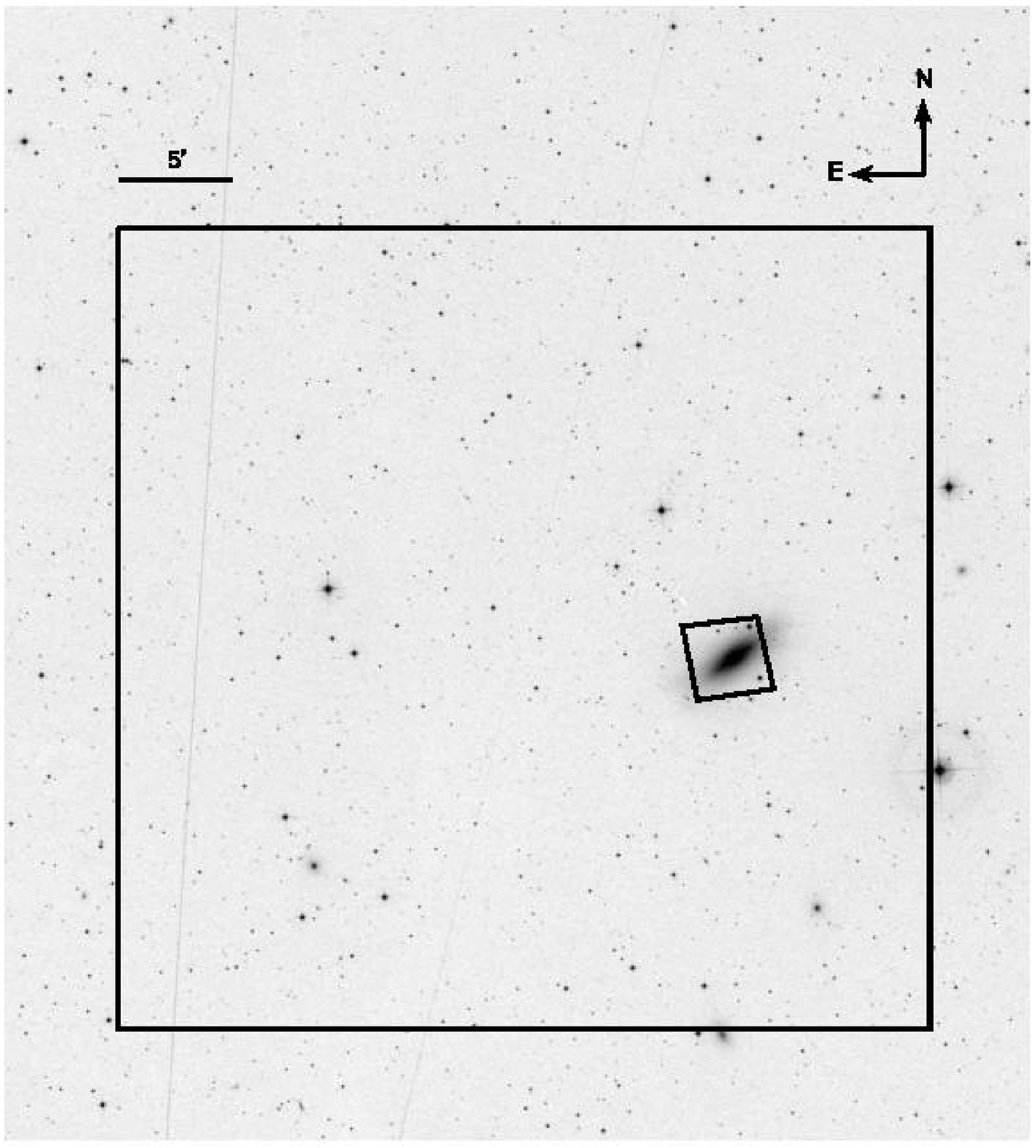}
\caption{Digitized Sky Survey image of NGC~5866 showing the area covered by the Mosaic imaging analyzed in this study (large box).  The {\it HST} ACS pointing analyzed in this study is also shown (smaller boxes).
\label{n5866_finder}}
\end{figure}

\begin{figure}
\plotone{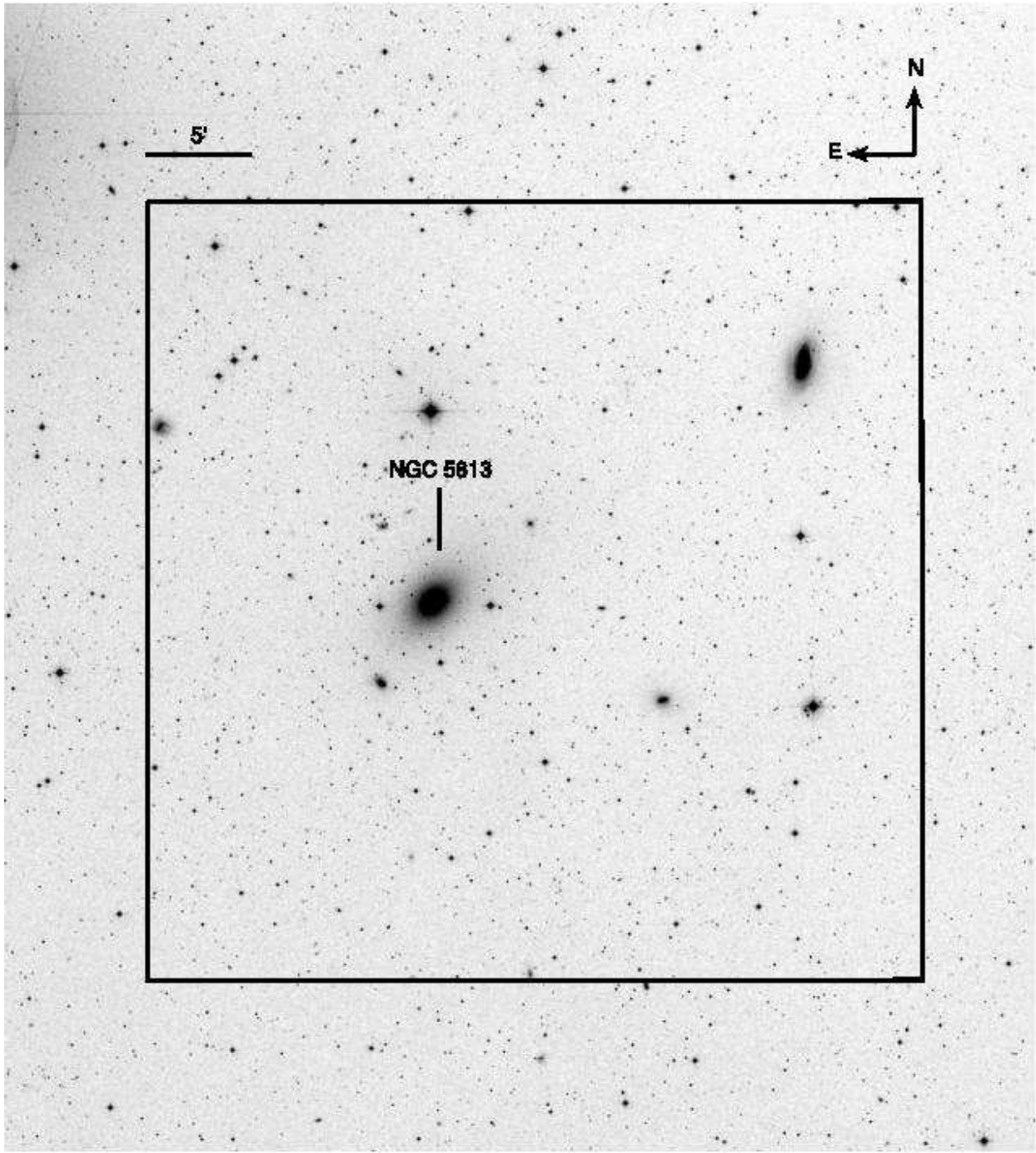}
\caption{Digitized Sky Survey image of NGC~5813 showing the area covered by the Mosaic imaging analyzed in this study (large box).
\label{n5813_finder}}
\end{figure}

\begin{figure}
\plotone{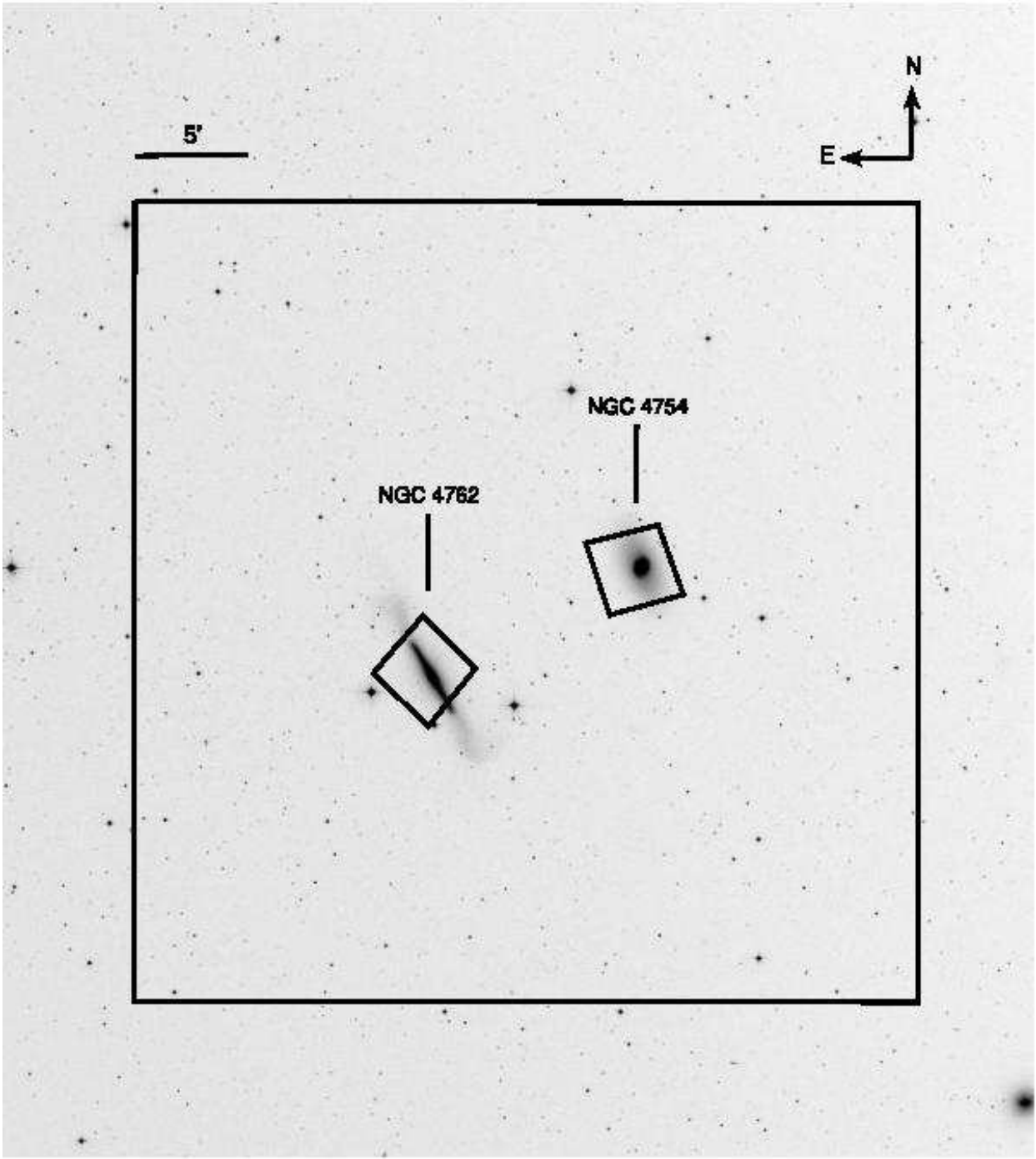}
\caption{Digitized Sky Survey image of the NGC~4754/4762 field showing the area covered by the Mosaic imaging analyzed in this study (large box).  The two {\it HST} ACS pointings analyzed in this study are also shown (smaller boxes).
\label{n4754_finder}}
\end{figure}

\begin{figure}
\plotone{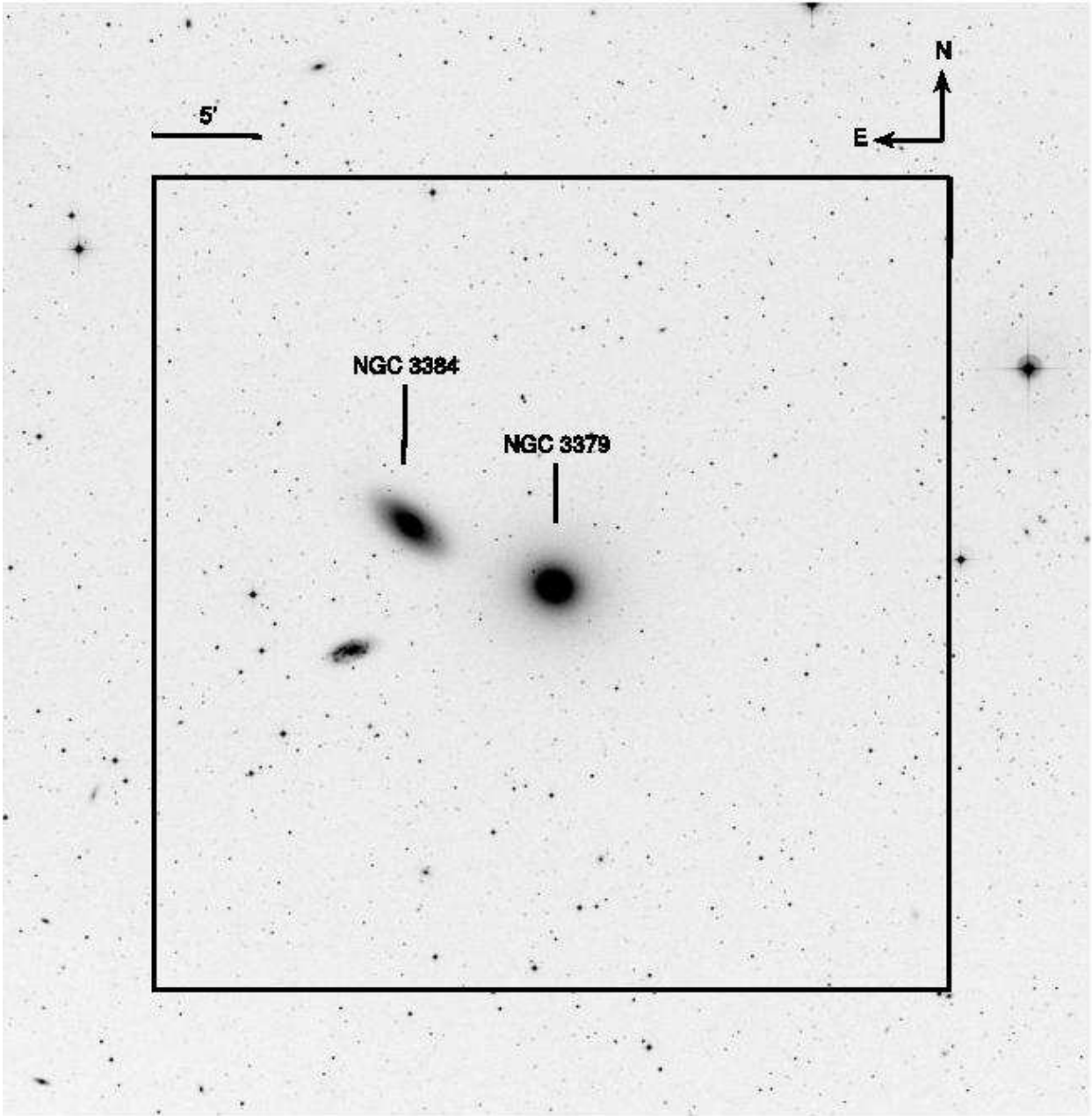}
\caption{Digitized Sky Survey image of the NGC~3384/3379 field showing the area covered by the Mosaic imaging analyzed in this study (large box).  
\label{n3384_finder}}
\end{figure}

The target galaxies were imaged using the Mosaic camera on the KPNO Mayall 4m telescope.  The Mosaic camera consists of eight $2048\times4096$ CCDs separated by a small gap ($\sim 50$ pixels).  On the Mayall telescope, the imager has a $36\arcmin \times 36\arcmin$ field of view with $0.26\arcsec$ pixels. The images of four of the target galaxies (NGC~5866, NGC~5813, NGC~4762, NGC~4754) were taken over five nights in May 2010.  NGC~3384 was imaged in March 1999 as part of a study of the GC systems of four E/S0 galaxies (RZ04).  For each galaxy, we obtained several images in each of three broadband filters ($BVR$), dithering between exposures to facilitate cosmic ray rejection and provide sky coverage across the CCD chip gaps. NGC~4762 and NGC~4754 are separated by $\sim 11\arcmin$ and were included a single Mosaic pointing.   The images of NGC~3379 from RZ04 also include NGC~3384 because the galaxies have an angular separation of $\sim 7\arcmin$.  We used the reduced images of the NGC~3379/3384 field from RZ04 for our analysis of NGC~3384.  Figures~\ref{n5866_finder}-\ref{n3384_finder} show the field of view of the Mosaic pointings for the target galaxies.

The observation dates, exposure times, and number of exposures used in our final, stacked image for each target are listed in Table~\ref{tbl_observations}. 
During the May 2010 observing run, night two was photometric. The remaining nights were non-photometric but generally clear with occasional clouds on some nights. In order to postcalibrate the data taken on the non-photometric nights, we obtained shorter single exposures of each target in each filter (exposure times of 600 s) and images of \citet{la92} standard fields on the photometric night.  Using these we derived a photometric calibration solution consisting of color coefficients and zero points.    
For the March 1999 observing run, the NGC~3379/3384 data were taken under non-photometric conditions and postcalibrated using additional observations as detailed in RZ04.  

The data were reduced using the MSCRED routines in IRAF \footnote{IRAF is distributed by the National Optical Astronomy  Observatory, which is operated by the Association of Universities for Research in Astronomy, (AURA), under cooperative agreement with the National Science Foundation.}.  The data were first processed using \texttt{ccdproc}, applying the overscan and bias level corrections. Flat fielding was done using master dome flats and median-smoothed master twilight flats.  The individual Mosaic extensions were then combined into a single FITS image using the \texttt{msczero}, \texttt{msccmatch}, and \texttt{mscimage} routines.  We next measured and subtracted a constant sky background level from each image.  Where smooth sky background gradients were present we used the \texttt{imsurfit} routine to fit and subtract a 2nd-order polynomial surface from the background, having first masked the chip gaps, stars, and target galaxies to obtain an accurate measure of the background variations.  Lastly, the series of images for each filter were scaled to a common flux level by measuring $\sim 20$ bright stars on each image. In each filter, the images for a particular target galaxy were aligned and a single stacked image was created using \texttt{mscstack} with \texttt{ccdclip} pixel rejection and no zero point offsets or additional scaling.  The mean background level of the image used as the flux scaling reference frame was added back to the stacked image.  The range of the mean full-width at half-maximum (FWHM) of the image point-spread functions (PSFs) for the final images is as follows: $1\farcs1$ to $1\farcs2$ for NGC~5866, $0\farcs9$ to $1\farcs2$ for NGC~5813, $1\farcs2$ to $1\farcs4$ for NGC~4762/4754, and $1\farcs3$ to $1\farcs4$ for NGC~3384.

\section{Detection of the Globular Cluster System}\label{sec_detection}

\subsection{Point Source Detection and Aperture Photometry}\label{sec_ptsource}

In order to detect the GC systems of the target galaxies, we use the procedures developed in our previous wide-field imaging studies.  We create a galaxy-subtracted image by smoothing the final combined image, subtracting it from the original, and restoring the constant sky background level.
All subsequent steps in the analysis were performed on these galaxy-subtracted images.  We detected all sources on the images using the \texttt{daofind} routine in IRAF with $4\sigma$ to $5\sigma$ detection thresholds in each of the three filters.  Regions of the image which contained saturated pixels (due to bleed trails from bright stars) or high noise levels (along the frame edges or near the centers of the host galaxy) were masked out.  The source lists in each filter were matched to proudce a final source list per field.  The total numbers of detected objects in $BVR$ are 3858 for NGC~5866, 10207 for NGC~5813, 4903 for NGC~4762/4754, and 3224 for the NGC~3384/3379 field.

\begin{figure}
\plotone{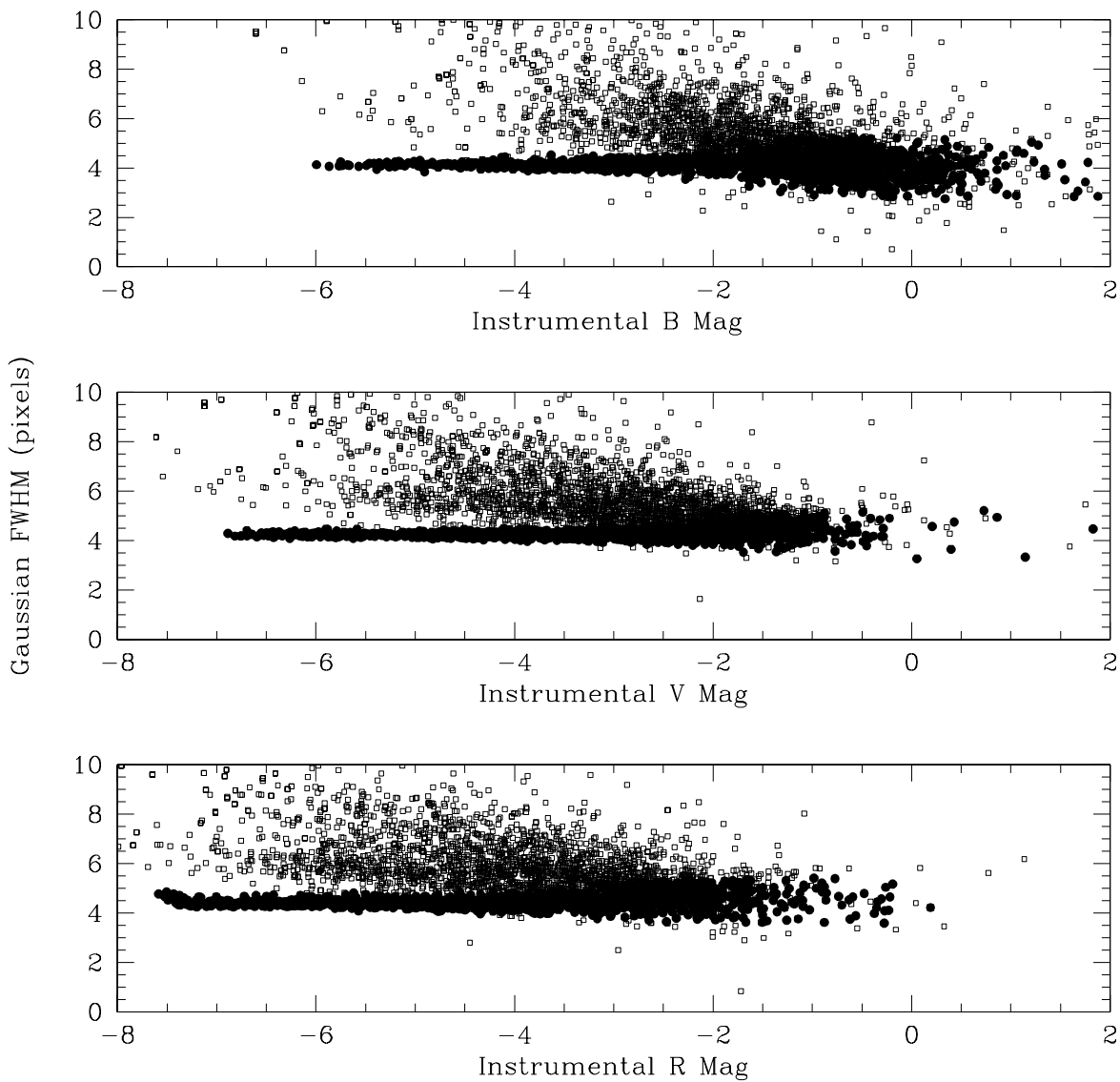}
\caption{Example point source selection for the detected objects in the Mosaic images.  Plotted are the full-width at half-maximum (FWHM) of the Gaussian profiles versus instrumental $BVR$ magnitudes of detected sources which were selected as point sources (filled circles) and rejected as extended objects (open squares). Extended source cuts similar to this example were done for all target galaxies.
\label{fig_escut}}
\end{figure}

Because GCs at the distances of the target galaxies will be point sources in our images,
we removed extended sources from our list of detected objects.  We used a graphical software routine to select the sequence of point sources in a plot of object FWHM versus instrumental magnitude, accounting for the increasing scatter at faint magnitudes.  Figure~\ref{fig_escut} shows an example of the point source selection for NGC~5866.  After requiring that objects pass the point source criteria in each of the $BVR$ filters, we are left with the following number of point sources for the target galaxies: 1551 for NGC~5866, 5709 for NGC~5813, 2167 for NGC~4762/4754, and 1728 for the NGC~3384/3379 field.  

Aperture photometry was performed on the point source objects in order to obtain calibrated magnitudes.  Here we describe the calibration procedures for the May 2010 Mosaic data. The photometry and calibration for the NGC~3384/3379 data was performed in a similar manner; the relevant details can be found in RZ04 Section 3.3.  We measured the point source objects using $1\times{\rm FWHM}$ apertures and applied aperture corrections to obtain total magnitudes.  The aperture corrections were determined using the mean difference between the total magnitude and the magnitude within an aperture of $1\times{\rm FWHM}$ for $15-20$ bright stars on the images.  The aperture corrections ranged from -0.443 to -0.222 with errors of 0.001 to 0.004.  Because the final $BVR$ science images and the photometric calibration data were taken on different nights (see Section \ref{sec_observations}), we determined bootstrap offsets to the instrumental magnitudes of the photometric night.  These offsets were determined using the final science images and the series of shorter exposure images taken on the photometric night.  The bootstrap offsets ranged from -0.406 to -0.007 with errors of 0.002 to 0.008.  The final calibrated magnitudes were calculated by applying the aperture corrections, atmospheric extinction corrections, and bootstrap offsets to the instrumental magnitudes determined using a photometric aperture with a radius of $1\times{\rm FWHM}$.  Finally, we corrected the calibrated magnitudes for Galactic extinction using the reddening maps of \citet{sfd98}.  Table~\ref{tbl_targets} lists $A_V$ for the target galaxies.  For the NGC~4762/4754 field, we used the extinction corrections for the center of the frame for both galaxies.  The difference in extinction across the field ranged from only 0.01-0.02 magnitudes, and therefore has a negligible impact on our selection of GC candidates.  

\subsection{Color Selection}\label{sec_colorcut}

Our list of GC candidates is determined by selecting objects from our point source list that have magnitudes and $BVR$ colors consistent with those of GCs at the distances of the target galaxies.  We adopt the same selection procedures used in our previous GC system studies and here only summarize the steps described in detail in RZ01.  We assume a range of GC absolute magnitudes based on observations of the Milky Way GC luminosity function (GCLF) and other well-studied nearby galaxies ($-11 \leq M_V \leq -4$; Ashman \& Zepf 1998).  We select as GC candidates those point sources that have $B-V$ and $V-R$ colors consistent with the Milky Way GCs but extend the range of allowed metallicities ([Fe/H] = $-2.5$ to $0.0$).  Lastly, we also consider the likelihood of a point source being a GC by examining the appearance of objects in {\it HST} imaging (see Section~\ref{sec_hst}) and their radial distance from the galaxy center.

\begin{figure}
\plotone{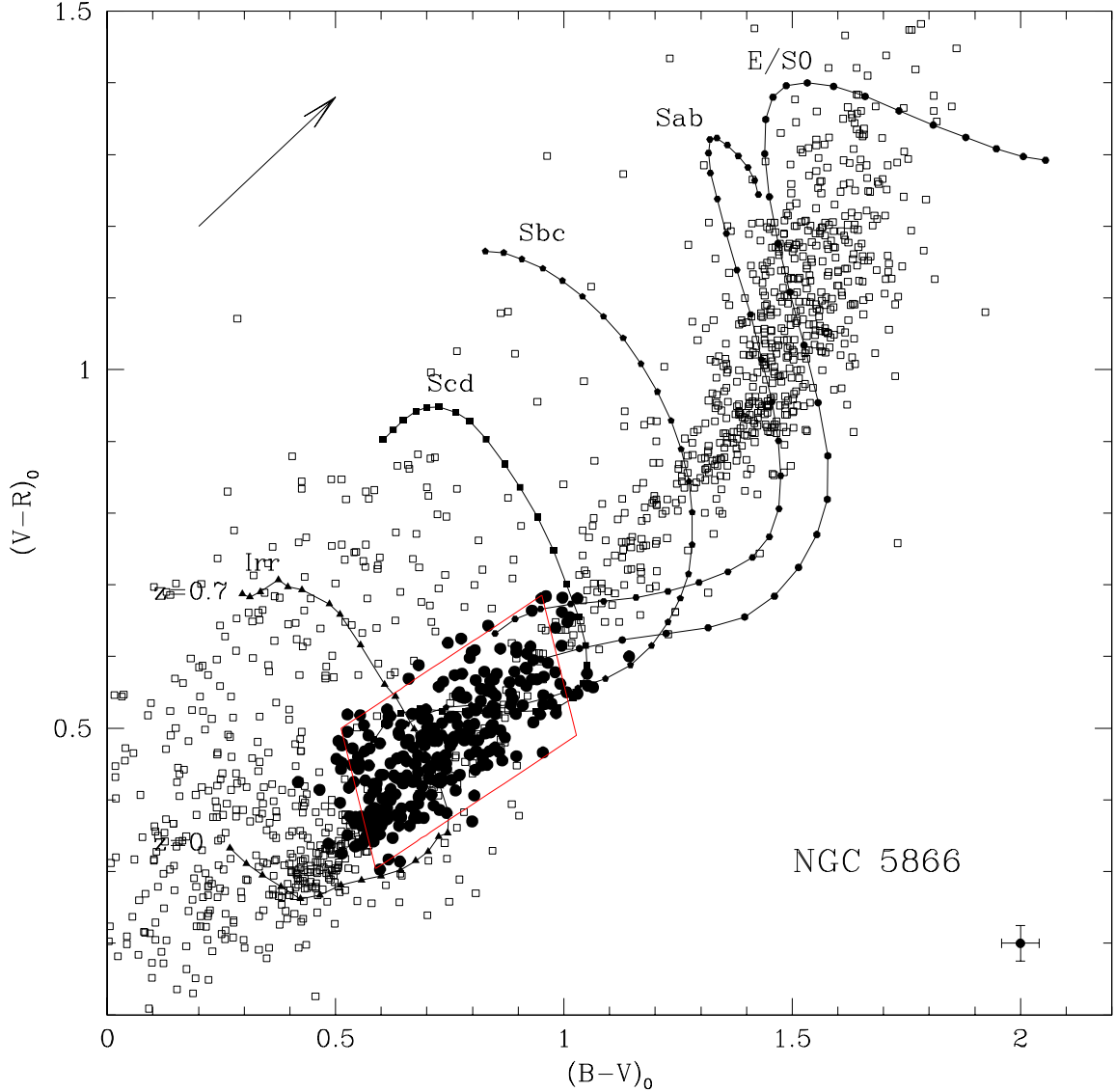}
\caption{Selection of the Mosaic GC candidates in NGC~5866 based on $V$ magnitudes and $B-V$, $V-R$ colors. The point sources detected in all three filters are shown as open squares and the final sample of 290 GC candidates are shown as filled circles. The red box denotes the boundary of the color selection (see Section~\ref{sec_colorcut}).  The imposed $V$ magnitude cut rejects some point sources in the selection box. Objects outside the selection box included as GC candidates results from the consideration of photometric errors in the selection process.  The lower right corner shows the median $B-V$ and $V-R$ photometric errors for the GC candidates. The curves denote the location of galaxies of various morphological types from redshift $z=0$ to $0.7$.  The upper left corner shows the reddening vector corresponding to $A_V = 1$.
\label{n5866_bvr}}
\end{figure}

\begin{figure}
\plotone{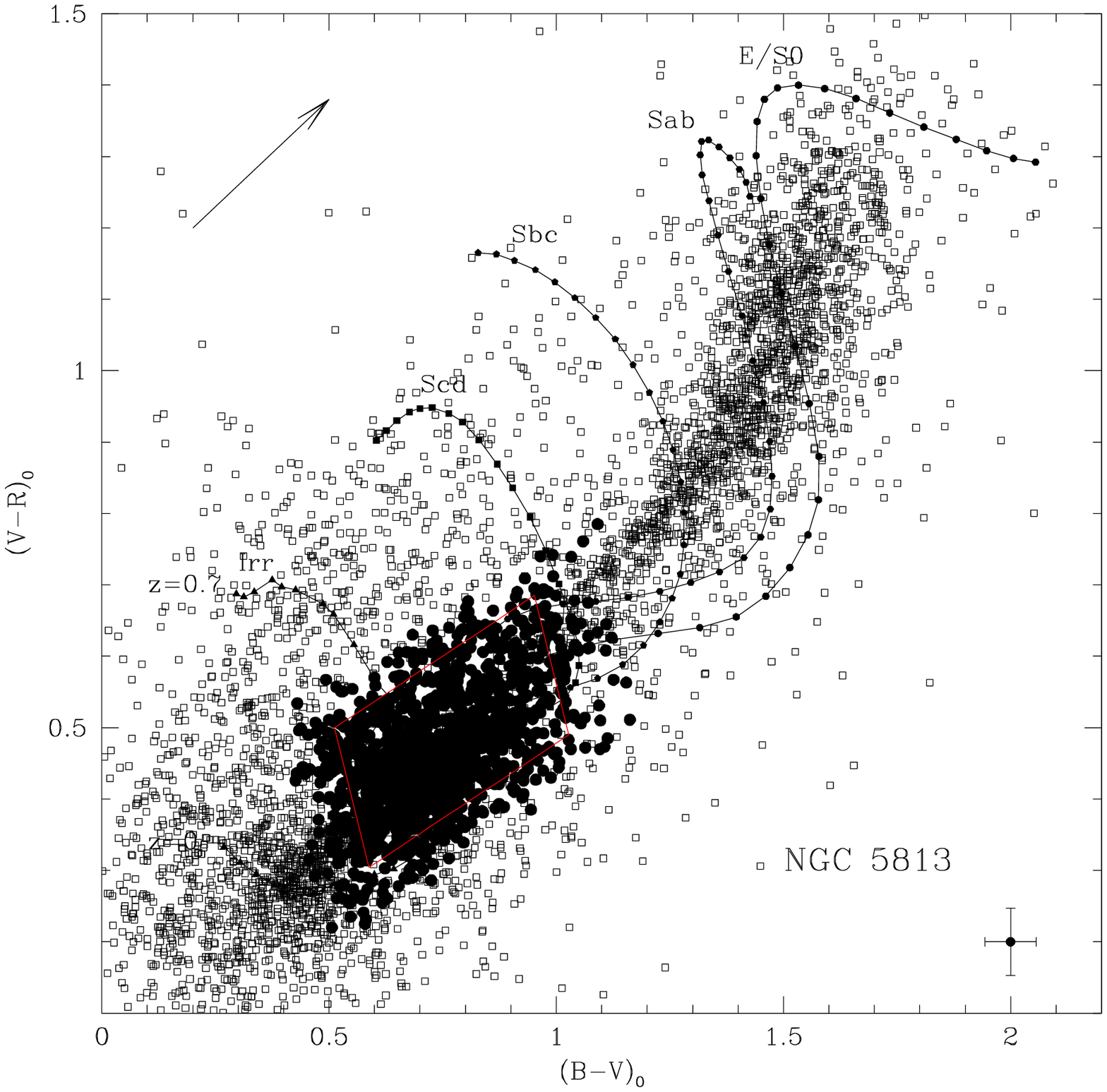}
\caption{Selection of the Mosaic GC candidates in NGC~5813 based on $V$ magnitudes and $B-V$, $V-R$ colors. 
The point sources detected in all three filters are shown as open squares and the final sample of 1300 GC candidates are shown as filled circles.
See caption for Figure~\ref{n5866_bvr} for additional plot details.
\label{n5813_bvr}}
\end{figure}

\begin{figure}
\plotone{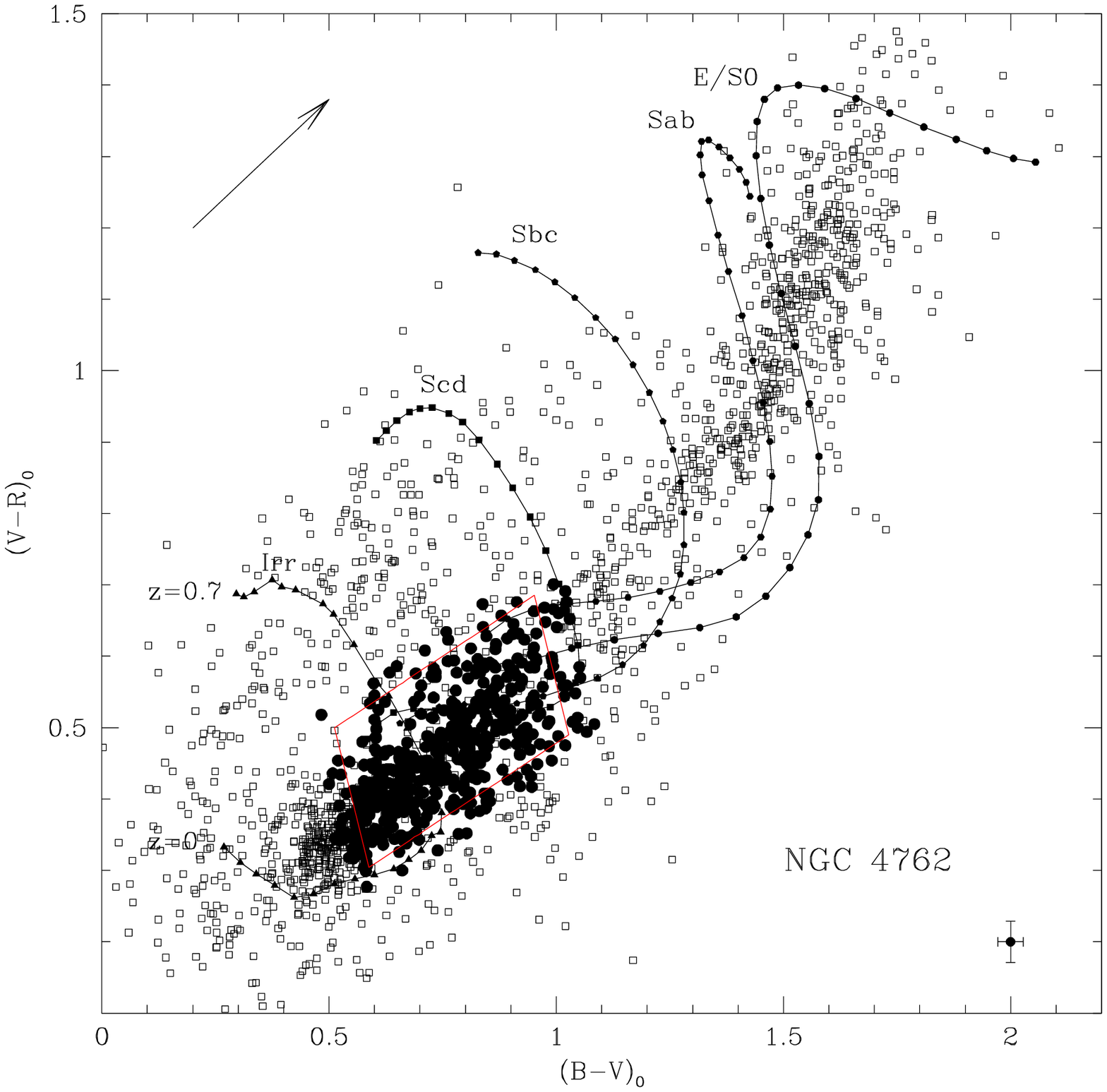}
\caption{Selection of the Mosaic GC candidates in NGC~4762 based on $V$ magnitudes and $B-V$, $V-R$ colors.  
The point sources detected in all three filters are shown as open squares and the final sample of 481 objects (GC candidates) with $V$ magnitude and $BVR$ colors that meet our selection criteria are shown as filled circles.  The selection has been done for the entire Mosaic image $(\sim 36\arcmin \times 36 \arcmin)$ with the region around NGC~4754 masked out.  See caption for Figure~\ref{n5866_bvr} for additional plot details.
\label{n4762_bvr}}
\end{figure}

\begin{figure}
\plotone{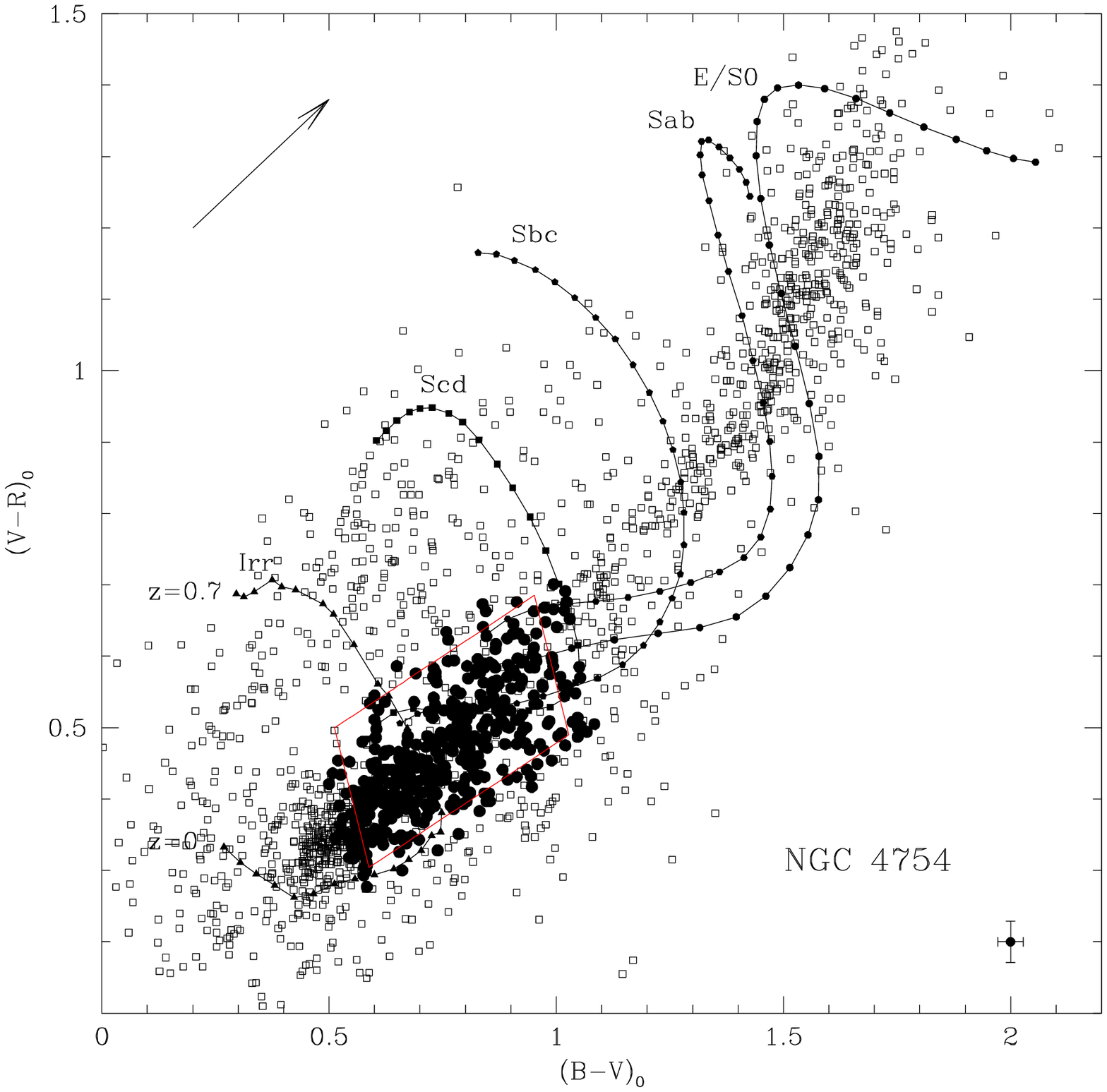}
\caption{Selection of the Mosaic GC candidates in NGC~4754 based on $V$ magnitudes and $B-V$, $V-R$ colors. 
The point sources detected in all three filters are shown as open squares and the final sample of 451 objects (GC candidates) with $V$ magnitude and $BVR$ colors that meet our selection criteria are shown as filled circles. The selection has been done for the entire Mosaic image $(\sim 36\arcmin \times 36 \arcmin)$ with the region around NGC~4762 masked out. See caption for Figure~\ref{n5866_bvr} for additional plot details.
\label{n4754_bvr}}
\end{figure}

\begin{figure}
\plotone{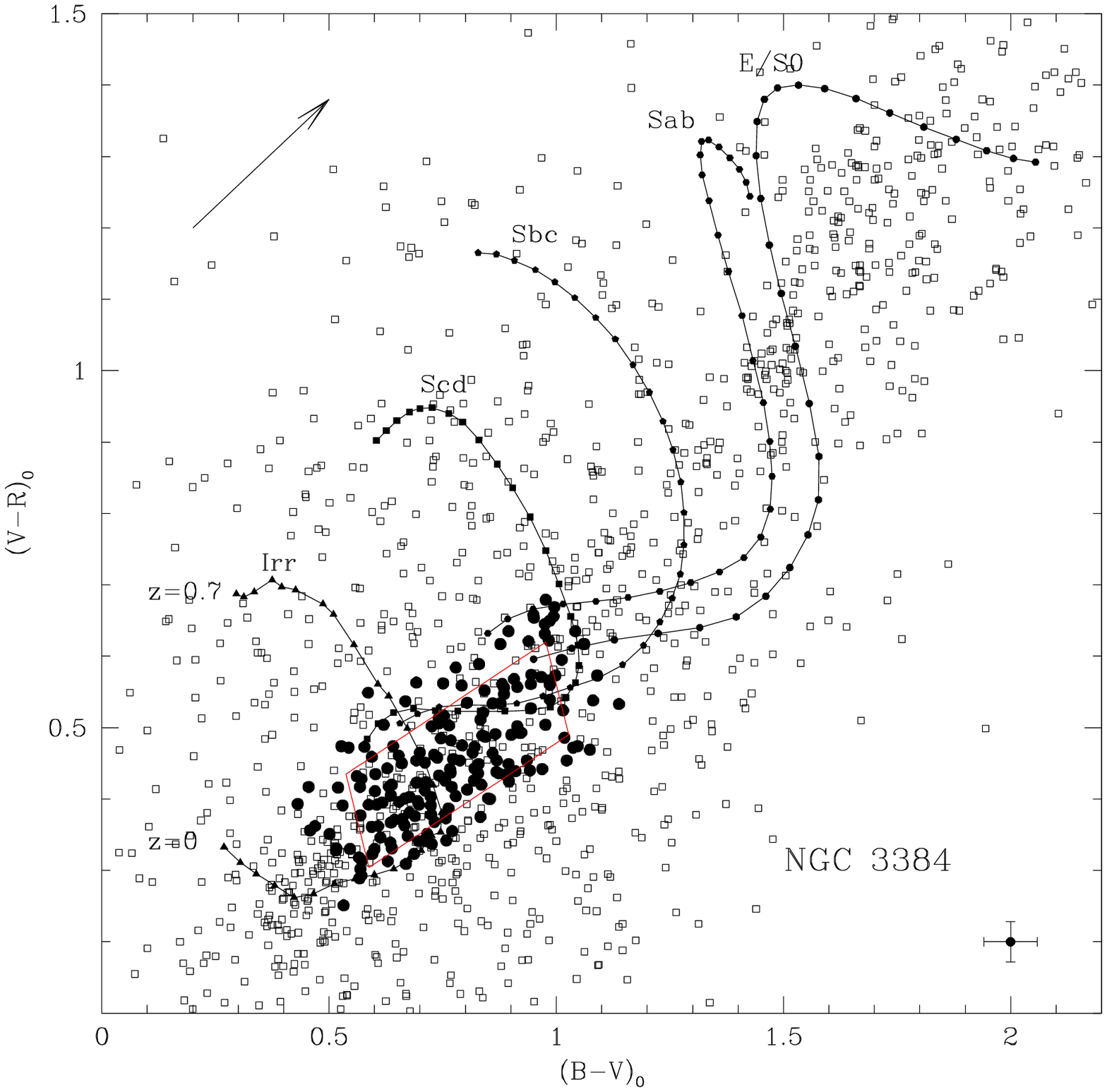}
\caption{Selection of the Mosaic GC candidates in NGC~4754 based on $V$ magnitudes and $B-V$, $V-R$ colors.  
The point sources detected in all three filters are shown as open squares and the final sample of 181 GC candidates are shown as filled circles.
See caption for Figure~\ref{n5866_bvr} for additional plot details.
\label{n3384_bvr}}
\end{figure}

Figures~\ref{n5866_bvr}-\ref{n3384_bvr} show the locations of the point sources in the $BVR$ color-color plane.  Because contamination from unresolved background galaxies is problematic in selecting GC candidates from ground-based imaging studies, we illustrate the expected colors of galaxies of various morphological types from $z=0.0$ to $0.7$ (see RZ01 for details).  The color selection process was performed for each galaxy as follows.  First, we removed any point sources brighter than the brightest expected GC at the distance of the galaxy ($M_V=-11$).  Second, because the faintest point sources will have large photometric errors, we applied a faint-end magnitude cut to keep only those GC candidates with magnitude errors in all filters $\lesssim 0.05$ mag (color errors $\lesssim 0.07$ mag).  This cut ensures we keep only the most well-measured GC candidates and, because the number of background galaxies grows at fainter magnitudes, this also reduces contamination.  Because such a cut could eliminate faint but {\it bona fide} GCs, we examined the spatial distribution of the objects being cut to ensure a lack of spatial concentration of objects around the galaxy center.  We note also that this magnitude cut is accounted for in our correction for magnitude incompleteness by the fitting of the GCLFs (see Section~\ref{sec_gclf}).  Third, we selected as GC candidates those remaining point sources which have colors (including photometric errors) within $3\sigma$ of the Milky Way $BVR$ color-color relation, expanded to include metallicities from [Fe/H] = $-2.5$ to $0.0$.  We examined in detail objects near the galaxy center which only barely fail the selection criteria, ensuring we did not miss objects which should be included in our final GC candidate lists.  Lastly, we removed any objects from our GC candidate list which are likely background galaxies as determined from our analysis of archival {\it HST} images (see Section~\ref{sec_background}). 

For the NGC~4762/4754 field, the color selection procedure required additional steps.  For NGC~4762, we masked a region around NGC~4754 in order to eliminate GC candidates likely associated with this galaxy.  We removed any GC candidates within a $4\arcmin$ radius of NGC~4754, having estimated the spatial extent of the galaxy's GC system using a radial surface density profile of GC candidates (see Section~\ref{sec_asymp}).  For NGC~4754 we applied the same masking procedures, removing all GC candidates within a $5\farcm7$ radius of NGC~4762 from the NGC~4754 GC candidate list.  Lastly, because the galaxies were imaged together, we adopted a single faint-end magnitude cut (based on the photometric errors) for both galaxies.  We used slightly different bright-end magnitude cuts due to the differing distances of the galaxies.  Figures~\ref{n4762_bvr} and~\ref{n4754_bvr} show the resulting color-color diagrams.  The magnitude and color cuts have been applied across the entire field; no spatial/radial cuts have been applied other than the respective galaxy masks. Therefore, because of the large field of view of the Mosaic images $(\sim 36\arcmin \times 36 \arcmin)$, the final GC candidate lists for both galaxies contained objects in common.  

For NGC~3384, the color selection followed the same procedure described for NGC~3379 in RZ04. The very shallow depth of the $B$ image means the detected objects in all three filters are preferentially blue. Thus the overdensity of GC candidates in the color-color plane was markedly skewed toward the blue side of the mean Milky Way $BVR$ GC relation.  To pick up all the objects in the overdensity, we used a $1\sigma$ width on the red side of the selection box and a $3\sigma$ width on the blue side of the box.  In addition, we masked the GC system of NGC~3379 in our determination of the NGC~3384 GC candidates.  We eliminated all GC candidates with a $6\arcmin$ radius of NGC~3379, adopting a mask size that extends to the radius where the final, corrected radial surface density profile is first consistent with zero within the errors (RZ04).  

After applying the color and magnitude selection steps, the final number of GC candidates for the target galaxies are as follows: 290 for NGC~5866, 1300 for NGC~5813, 481 for NGC~4762, 451 for NGC~4754, and 181 for NGC~3384.

\section{Analysis of the Globular Cluster System}\label{sec_analysis}

\subsection{Completeness Testing and Detection Limits}\label{sec_complete}

The point source detection and magnitude completeness limits were measured using a series of artificial star tests on the $BVR$ sciences images.  We first computed a model PSF for each image derived from the best-fit average PSF of the frame.  In a single test, we added 600 artificial stars to the image with magnitudes which vary around 0.2 mag of a specific value.  We next applied the identical point source detection criteria used to detect each galaxy's GC system (see Section~\ref{sec_ptsource}) and computed the fraction of detected artificial stars.  We ran the tests over a 4-5 magnitude range and thus obtained a measure of the completeness (fraction of detected point sources) as a function of magnitude in each filter.  Table~\ref{tbl_completeness} lists the 90\% and 50\% completeness limits for the galaxies in this study.

\subsection{Contamination Corrections}\label{sec_contamination}

The GC system detection process (see Section~\ref{sec_detection}) significantly reduces contamination in our GC candidate lists from bright Galactic foreground stars, spatially-extended background galaxies, and objects with colors inconsistent with those of confirmed GCs and their range of metallicities. Even so, our GC candidate lists contain some degree of contamination.  Below we discuss how we quantify the remaining contamination in the GC candidate samples.  

\subsubsection{Contamination Estimates based on the Asymptotic
  Behavior of the Radial Profile}\label{sec_asymp}

The large spatial coverage obtained with the Mosaic camera provides a means of directly quantifying the level of contamination without additional pointings and/or control fields.  Wide-field imaging studies of extragalactic GC systems typically show a centrally-concentrated profile when plotting the azimuthally averaged surface density of GC candidates as a function of projected radius from the host galaxy center.  Assuming the spatial profile of the GC system is not larger than the imaging area, the radial profile decreases smoothly to a mean background level.  Because at large projected radii very few {\it bona fide} GC will be present (as evidenced by the decreasing radial profile), the population will be dominated by contaminating foreground Galactic stars and background galaxies.  Measuring the mean surface density (over a large area) at large projected radii from the host galaxy center thus provides a direct empirical estimate of our sample contamination.  The radial surface density profiles can then be corrected for contamination by subtracting the measured asymptotic surface density of contaminants.

We constructed radial surface density profiles for the target galaxies and measured the asymptotic contamination values as detailed below.  In all cases we used circular annuli and explored the effects of choices of bins sizes and bin center locations on the shape of the radial profile. We accounted for missing area in the bins (due to masked regions or bins which extend off the images) by using the ``effective area", i.e., calculating the area of unmasked pixels.  After adopting a contamination estimate, we computed the radially-dependent contamination fraction as the ratio of the number of contaminating objects in the bin (surface density of contaminants multiplied by the effective area of the bin) to the total number of GC candidates in the bin.  We describe the initial radial profiles and asymptotic contamination estimates for each galaxy below.

{\it NGC~5866}. -- We constructed the radial profile using bins with $2\farcm5$ spacing.  Given the relatively small number of GC candidates for this galaxy ($N=290$) spread over the entire Mosaic frame, our experiments with various radial bin widths indicated that wider bins were necessary to ensure a significant number of GC candidates per bin and produce a smooth radial profile. The region inward of $\sim 1\arcmin$ was masked due to the poorer quality of the $B$ band imaging, so the bin centers ranged from $1\arcmin$ from the galaxy center to $28\farcm5$.  The surface density falls to a nearly constant level in the bins outward of $\sim10\arcmin$, indicating that we have observed the full radial extent of the GC system.  We computed the asymptotic contamination correction as the weighted mean of the surface density in the five radial bins from $r=14\farcm8-24\farcm7$, finding a value of $0.21 \pm 0.02~{\rm arcmin}^{-2}$ (assuming Poisson statistics).

{\it NGC~5813}. -- We constructed the radial profile using bins with $1\arcmin$ spacing.  The region inward of $\sim0\farcm4$ is masked, so the radial bins ranged from $0\farcm4$ from the galaxy center to $26.4\arcmin$.  The surface density decreases smoothly to a nearly constant level starting at the $r=12.9\arcmin$ bin but shows some small fluctuations between $13\arcmin$ and $18\arcmin$.  We calculated the asymptotic surface density as the weighted mean of the surface density in the ten bins between 15\arcmin and 24\arcmin, finding a value of $0.69 \pm 0.04~{\rm arcmin}^{-2}$ (assuming Poisson statistics).  

{\it NGC~4762}. -- We constructed the radial profile using bins with $1\farcm25$ spacing ranging from $0\farcm5$ to $28\arcmin$.  The surface density decreased smoothly from the galaxy center, falling to a nearly constant level at the $\sim 4\farcm9$ bin.  To avoid possible contribution from the NGC~4754 GC system, we measured the contamination using the area outside the GC systems of both galaxies.  Using a single large radial bin with an inner edge at $15\arcmin$ and extending to $28\arcmin$, we find a surface density of contaminants of $0.40 \pm 0.03~{\rm arcmin}^{-2}$ (assuming Poisson statistics).  

{\it NGC~4754}. -- We constructed the radial profile using bins ranging from $0\farcm45$ to $26\farcm7$ using $0\farcm75$ wide radial bins.  The profile decreased smoothly to a nearly constant level at the $3\farcm1$ radial bin. 
Because the GC candidate list for NGC~4754 differs from NGC~4762 (due to the different bright-end magnitude cuts), we cannot use the NGC~4762 contamination estimate.  
We avoid any contribution from the NGC~4762 GC system by using a single large radial bin outside the radial extent of both GC systems.  Using one bin with an inner edge at $17\arcmin$ and extending to $28\arcmin$, we find a surface density of contaminants of $0.40 \pm 0.03~{\rm arcmin}^{-2}$ (assuming Poisson statistics), consistent with the contamination estimate for NGC~4762.

{\it NGC~3384}. -- We constructed the initial radial profile using bins with $1\arcmin$ spacing ranging from $0\farcm59$ (inner edge of innermost bin) to $18\farcm59$ (outer edge of outermost bin).  The profile shows a smooth decrease to a nearly constant level at the $r=5\farcm1$ radial bin.  The surface density profile showed a very slight but insignificant increase above the errors in the $r=8\farcm1$ radial bin.  We calculated an asymptotic contamination surface density of $0.18\pm0.02$ (assuming Poisson statistics) using the nine radial bins from $10\farcm1$ to $18\farcm1$, adopting the weighted mean of the surface density in these bins as our final value.  This value is in good agreement with the asymptotic contamination surface density estimate for NGC~3379 of $0.22\pm 0.02~{\rm arcmin}^{-2}$ from RZ04.

\subsubsection{Models of Foreground Galactic Contamination}\label{sec_foreground}

A primary source of contamination in our wide-field imaging studies are foreground Galactic stars (predominately in the halo) that have magnitudes and colors that pass the GC selection criteria.  We can estimate the degree of this contamination and compare the results to our measured asymptotic contamination levels using Galactic star count models.  We used the Besan\c{c}on Galactic stellar population synthesis models of \citet{ro03} to predict the number counts of Milky Way stars in our observed fields.  Briefly, the Besan\c{c}on models produce a four component structural model of the Galaxy (thin disk, thick disk, spheroid, bulge) using stellar evolution tracks, initial mass functions (IMFs), age ranges, and metallicities for each stellar population.  The simulations output a catalog of stars which pass the user-defined magnitude range(s), color range(s), and area on the sky.  

\begin{figure}
\plotone{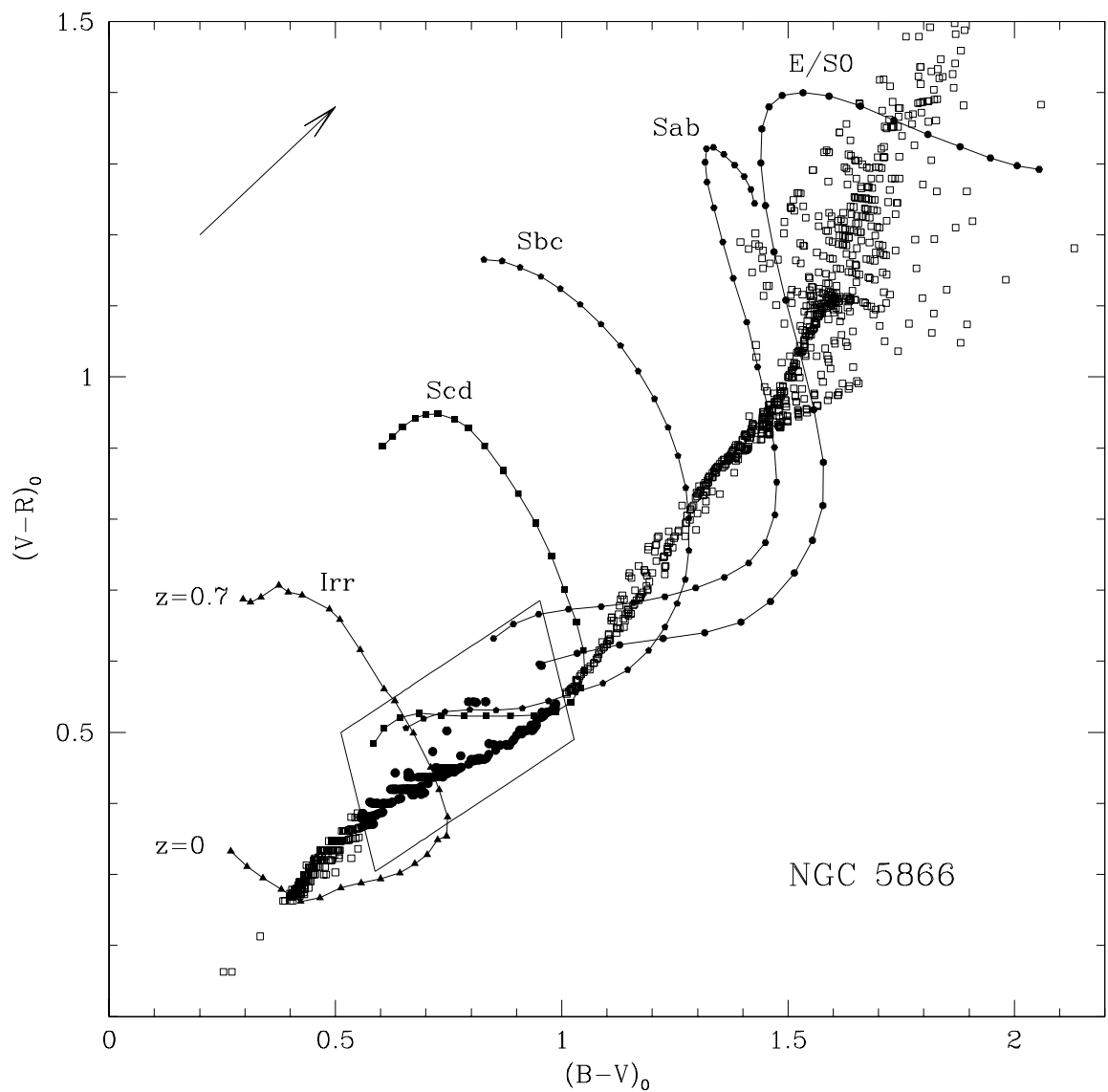}
\caption{Color-color diagram for a Besan\c{c}on Galactic star count model for the NGC~5866 Mosaic field.  Estimates of the foreground Galactic contamination were made using the Besan\c{c}on stellar population synthesis models (see Section~\ref{sec_foreground}).  The $V$ magnitude range of the model stars was chosen to match the GC candidate selection range for NGC~5866 (see Section~\ref{sec_colorcut}). Open squares show the $B-V$ and $V-R$ colors of all model stars.  The scatter in the colors is intrinsic to the simulations.
Solid circles show the model stars that would pass the color selection criteria (defined by the box) for our GC candidates and would be considered contaminating objects.  The background galaxy curves and reddening vector are identical to Figures~\ref{n5866_bvr}-\ref{n3384_bvr} and are shown for comparison to these diagrams.
\label{besancon_galactic}}
\end{figure}

For each galaxy, we use the Besan\c{c}on models estimate the surface density of Galactic stars which pass our bright and faint $V$ magnitude cuts, $B-V$ and $V-R$ color cuts, and cover the identical area of sky of our Mosaic fields.  Because the models employ Monte Carlo methods to generate the stellar populations, we run the simulations ten times for each galaxy and use the mean and standard deviation of the ten runs to estimate the total number of contaminating stars.  We show an example simulation and color selection for NGC~5866 in Figure~\ref{besancon_galactic}, plotting the $BVR$ colors of simulated stars which fall in the magnitude range of the NGC~5866 GC candidates over the area of the Mosaic pointing.  Comparing the simulation to the NGC~5866 observations (Figure~\ref{n5866_bvr}) shows the effect of photometric uncertainties in our observations.  

The Besan\c{c}on estimates of the foreground Galactic contamination for the six target galaxies are as follows: $0.15~{\rm arcmin}^{-2}$ for NGC~5866, $0.29~{\rm arcmin}^{-2}$ for NGC~5813, $0.12~{\rm arcmin}^{-2}$ for NGC~4762, $0.12~{\rm arcmin}^{-2}$ for NGC~4754, and $0.11~{\rm arcmin}^{-2}$ for NGC~3384.  The typical uncertainty (standard deviation) in the surface density was $0.01~{\rm arcmin}^{-2}$. 
For all galaxies, the simulated Galactic contamination estimates are lower than the values derived from observations (using the asymptotic behavior of the radial profile; see Section~\ref{sec_asymp}). This is consistent with our expectation that the additional source of contamination comes from unresolved background galaxies.

\subsubsection{Estimates of Background Galaxy Contamination from \textit{HST} imaging of Mosaic GC Candidates}\label{sec_background}

Because of the superior spatial resolution provided by {\it HST} relative to our ground-based imaging, GC candidates that are unresolved in Mosaic data may be resolved as background galaxies when imaged by {\it HST}.  We can therefore obtain an estimate of the level of background galaxy contamination by examining GC candidates in archival {\it HST} images.  We downloaded Wide Field Planetary Camera 2 (WFPC2) and ACS/WFC data for the target galaxies from the Hubble Legacy Archive (HLA) \footnote{Based on observations made with the NASA/ESA Hubble Space Telescope, and obtained from the Hubble Legacy Archive, which is a collaboration between the Space Telescope Science Institute (STScI/NASA), the Space Telescope European Coordinating Facility (ST-ECF/ESA) and the Canadian Astronomy Data Centre (CADC/NRC/CSA).}, selecting data which were pipeline-reduced, calibrated, and combined to produce a single stacked image in each filter (Level 2 data).  Table~\ref{tbl_hst} lists the proposal ID, the target galaxy for the observations, the total exposure time of the combined images, the {\it HST} instrument, and filter for the archival {\it HST} data analyzed in this study.  After locating the Mosaic GC candidates on the {\it HST} images, we followed the method of \citet{ku99} and performed aperture photometry of the GC candidates using 0.5 pixel and 3 pixel apertures.  The ratio of counts in these apertures (larger aperture divided by smaller aperture) gives a measure of the degree of concentration of the light profile.  Relative to compact GCs, background galaxies will have a slightly larger spatial extent and therefore show a higher count ratio than GCs. 
Following \citet{ku99}, for the WFPC observations we use a threshold ratio of $> 8$ for objects imaged on the WFPC2 WF chips; we had no GC candidates imaged on the WFPC2 PC chip.  For the ACS/WFC images, we used a threshold ratio of $>12$.  We examine each flagged GC candidate visually in the {\it HST} images to verify that they have an extended spatial appearance.  As noted in Section~\ref{sec_colorcut}, we remove any probable background galaxies from our GC candidate lists.  Typically only 1-3 objects were confirmed background galaxies and therefore removing them from the GC candidate lists has a negligible impact on our final results.  In addition, because the {\it HST} pointings are centered on the galaxy, removing these objects does not affect the asymptotic contamination estimates determined in Section~\ref{sec_asymp}.

Using the number of background galaxies and the area of the {\it HST} pointing, 
we can obtain a measure of the surface density of contaminating background galaxies in the Mosaic GC candidate lists.  For NGC~5866, because the central regions of the Mosaic images were heavily masked, only six GC candidates were located on the ACS/WFC images.  Of these six, none of the GC candidates showed a high count ratio and therefore we obtained no estimate of the background galaxy contamination. For NGC~5813, we found that three of 55 Mosaic GC candidates are galaxies and we find surface density of $0.36~{\rm arcmin}^{-2}$. For NGC~4762, we found that only one of the 19 GC candidates was a background galaxy from which we find a surface density of $0.12~{\rm arcmin}^{-2}$.  For NGC~4754, found that three of the 16 GC candidates are background galaxies and we find a surface density of $0.31~{\rm arcmin}^{-2}$.  For NGC~3384, we adopted a background galaxy contamination estimate of $0.08~{\rm arcmin}^{-2}$ from the RZ04 study of NGC~3379.  RZ04 analyzed four {\it HST}/WFPC2 fields of the NGC~3379/3384 region, including a pointing centered on NGC~3384 (see Table~5 in RZ04). Because our selection of GC candidates around NGC~3384 is identical to the NGC~3379 study, the {\it HST} contamination analysis is also identical and provides a contamination estimate for our study of NGC~3384.  

In general we find that the background galaxy contamination estimates are lower than our empirical asymptotic contamination estimates.  This is consistent with our expectation that the remaining contamination comes from foreground Galactic stars.

\subsection{Coverage of the GCLF}\label{sec_gclf}

For each target galaxy we constructed the observed GCLF (corrected for contamination and magnitude completeness) following the procedure outlined in RZ01.  We assign the GC candidates to $V$ band magnitude bins of a specified width (bin widths of $0.5$ mag for NGC~3384; $0.4$ mag for NGC~5866, NGC~5813; $0.3$ mag for NGC~4762, NGC~4754).  We account for contamination by using the radially-dependent contamination fractions calculated in Section~\ref{sec_asymp}, applying these fractions as a correction factor.  Magnitude incompleteness at a given $V$ band magnitude is calculated using the artificial star test results (Section~\ref{sec_complete}) and the range of $B-V$ and $V-R$ colors of the GC candidates.  Following RZ01, we use the $BVR$ completeness results to derive a total (convolved) completeness which accounts for the individual completeness in all three filters.  The final, corrected GCLF is derived by dividing the total number of GC candidates in each magnitude bin (corrected for contamination) by the convolved completeness fraction at that magnitude.

\begin{figure}
\plotone{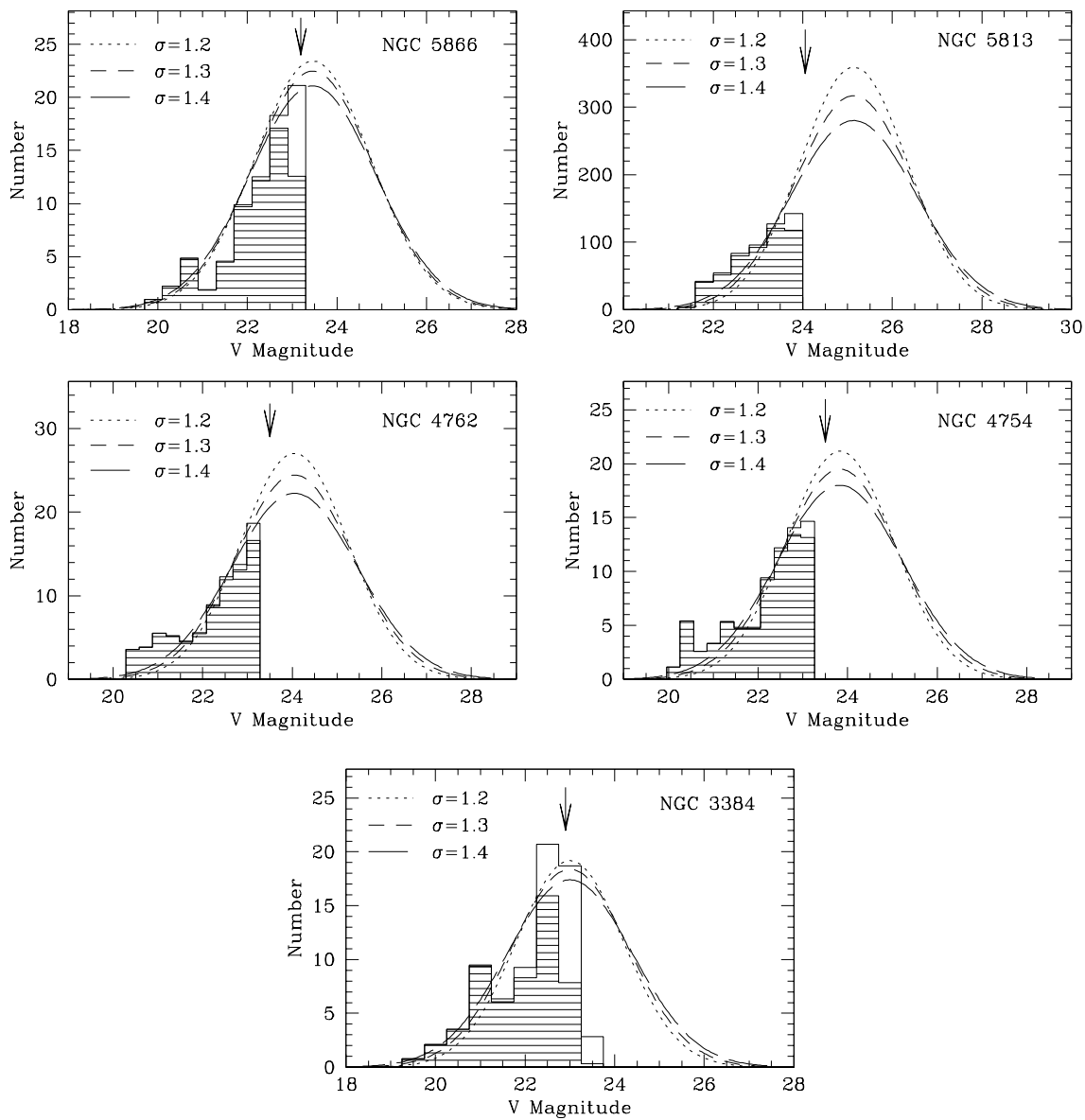}
\caption{GCLF fits for the target galaxies.  The observed GCLF is shown as the shaded histogram and the completeness-corrected GCLF (used in the fits) is shown as the solid line histogram.  The best-fit theoretical Gaussian functions are shown for three values of the dispersion: $\sigma=1.2$ (dotted line), 1.3 (short dashed line), and 1.4 (long dash).  The arrow denotes the magnitude where the convolved $BVR$ completeness is $50\%$.
\label{group_gclfs}}
\end{figure}

To determine the observed coverage of the theoretical GCLF, we assumed a Gaussian functional form for the theoretical GCLF and peak absolute magnitudes of $M_V = -7.3$ for S0/spiral galaxies and $M_V=-7.4$ for elliptical galaxies \citep{as98,ku01a}.  For the assumed distances of the target galaxies (see Table~\ref{tbl_targets}), the peak apparent $V$ magnitudes of the observed GCLFs are $V=25.1, 24.1, 23.8, 22.8$ for NGC~5813, NGC~4762, NGC~4754, and NGC~3384, respectively.  We adopted dispersions for the Gaussian fits of $1.2$, $1.3$, and $1.4$ mag, excluding bins from the fit which had very low numbers of GCs or were less than 70\% complete.  The observed and corrected GCLFs with theoretical fits are shown in Figure~\ref{group_gclfs}.  The mean fractional coverage of the theoretical GCLF by the observed GCLF for the target galaxies were as follows: $0.196 \pm 0.005$ for NGC~5813,  $0.313 \pm 0.002$ for NGC~4762, $0.351 \pm 0.001$ for NGC~4754, and $0.438 \pm 0.007$ for NGC~3384.  For the four galaxies, varying the choice of magnitude bin sizes and starting bin of the GCLF showed a $6\%-8\%$ change in the fractional coverage.  We account for this variation in our error estimates on the total number of GCs in Section~\ref{sec_numbers}.

For NGC~5866, we adopted the observed peak apparent magnitude of $V=23.46 \pm 0.06$ and dispersion of $\sigma = 1.13 \pm 0.05$ as measured by CBR07.  In their {\it HST} ACS study of the GC system of NGC~5866, CBR07 obtained nearly full coverage of the theoretical GCLF at better than 50\% completeness, yielding a robust measure of the peak apparent magnitude and dispersion of the GCLF.  Assuming our adopted distance to NGC~5866 we would expect a peak apparent magnitude of $V=23.6 \pm 0.1$, which agrees within the errors with the empirical measurement of CBR07. We fit the corrected, observed GCLF assuming three different $V$ magnitude bin widths ($0.3$, $0.4$, $0.5$ mag) and fixed values for the peak apparent magnitude and dispersion.  The mean fractional coverage of the theoretical GCLF by the observed GCLF was $0.409 \pm 0.004$ for NGC~5866.

\subsection{Analysis of Archival \textit{HST} Observations}\label{sec_hst}

In order to provide improved spatial coverage of the GC systems of the target galaxies, we complement our Mosaic imaging with archival and published {\it HST} imaging studies.  The superior spatial resolution of {\it HST} allows us to study the GC population closer to the galaxy center than the Mosaic imaging, providing data in a region where ground-based observations are often limited.  However, because the {\it HST} spatial coverage is relatively small, these data alone are often insufficient to measure the full radial extent of giant galaxy GC systems (RZ03; R10; H11). 
In this section we describe our analysis of archival and published {\it HST} ACS data for NGC~5866, NGC~4762, and NGC~4754.  The field of view of the ACS/WFC observations are shown in Figures~\ref{n5866_finder} (NGC~5866) and ~\ref{n4754_finder} (NGC~4762, NGC~4754).  For each pointing, the galaxies were offset only slightly from the center of the ACS/WFC field of view; the observations provide radial coverage out to $\sim 2\farcm7$ from the galaxy centers.  

{\it NGC~5866}. -- CBR07 studied the GC system of NGC~5866 using deep {\it HST} ACS/WFC observations in the F435W, F555W, and F625W filters (see Table~\ref{tbl_hst}).    
CBR07 selected GC candidates using size, shape, and color information for their detected sources, ultimately deriving a list of 109 GC candidates. 
The ACS filter magnitudes were converted to $BVR$ magnitudes using the transformations of \citet{si05}.  Comparing the CBR07 color selection criteria to our Mosaic color selection procedure showed excellent agreement and thus no additional color cuts were made to the CBR07 sample.  To account for magnitude incompleteness, we apply a faint-end magnitude cut at the $\sim 100\%$ completeness level ($V=23.9$) based on the completeness testing results described in CBR07.  
After applying the magnitude cut, we are left with 71 GC candidates which we use in subsequent steps in the analysis.

Because the expected contamination from unresolved background galaxies brighter than $V \sim 24$ should be relatively small in {\it HST} studies \citep{ku99}, we assumed no contamination when constructing the radial profile and GCLF for the NGC~5866 {\it HST} data.  CBR07 estimated a contamination level in their sample of only $6\%$ of the total number of GC candidates over the entire magnitude range.  Thus our sub-sample of 71 {\it HST} GC candidates, selected to be brighter than $V=23.9$, should have less than $6\%$ overall contamination.  Using our GC candidate list, we estimated the fractional coverage of the GCLF by the {\it HST} data in a manner similar to our Mosaic GCLF analyses.  The GCLF was fit using a Gaussian with a peak apparent magnitude of $V=23.46$ and dispersion of $\sigma=1.1$ as determined by CBR07.  We varied the magnitude bin sizes from $0.4$ to $0.6$ mag and find a mean fractional coverage of $0.69 \pm 0.01$.  Varying the choice of dispersion between $\sigma=1.0$ and $\sigma=1.2$ yielded a $2\%$ change in the fractional coverage.  We account for the variation in our error estimates on the total number of GCs in Section~\ref{sec_numbers}.  For the construction of the radial surface density profile (see Section~\ref{sec_radial}), we 
adopted circular radial bins with a $0\farcm4$ width ranging from $0\farcm17$ to $2\farcm3$.

{\it NGC~4762 and NGC~4754}. -- NGC~4762 and NGC~4754 were included as target galaxies in the ACS Virgo Cluster Survey \citep{co04} and imaged in the F475W and F850LP filters (Sloan Digital Sky Survey $g$ and $i$, respectively; see Table~\ref{tbl_hst}). 
Our GC candidate lists for NGC~4762 and 4754 were taken from the published catalogs of Jord{\'a}n et al. 2009 (their Table 4), who use magnitude, size, and shape criteria to derive catalogs of GC candidates for each ACSVCS galaxy. 
To account for magnitude incompleteness, we apply a faint-end magnitude cut at the $\sim 100\%$ completeness level based on the completeness tests described in \citet{jo09}.  The VCS study considers completeness as a function of object size, background surface brightness, and magnitude.  We computed the mean background surface brightness and mean size for the full VCS GC candidate lists and used $g$ band completeness curve (Table 2 from Jord{\'a}n et al. 2009) which corresponds closest to the mean values to determine the magnitude at which the completeness falls below $100\%$.  For both galaxies the full VCS GC candidate lists were cut at  $g=24.95$, leaving a total of $95$ and $69$ {\it HST} GC candidates for NGC~4762 and NGC~4754, respectively.  No magnitude cut was necessary at the brightest end of the GC candidate lists (to reject foreground Galactic stars), as the VCS GC candidate selections were already consistent with rejecting objects brighter than $M_V \lesssim -11$ at the adopted distances of the galaxies.  

To account for contamination, we used the VCS control fields to estimate the surface density of contaminants in the NGC~4762 and NGC~4754 fields, respectively.  The VCS uses the control fields to estimate contamination by scaling the control field observations to match the observations of each VCS target galaxy, analyzing the fields in a manner identical to the the GC candidate analysis, and deriving catalogs of point sources and their probabilities of being GCs.  After applying our GC candidate selection criteria 
to the control field catalogs, we find a total of $81$ and $89$ contaminating objects which, over 17 {\it HST}/ACS pointings, yields a surface density of contaminants of $0.42 \pm 0.05~{\rm arcmin}^{-2}$ and $0.46 \pm 0.05~{\rm arcmin}^{-2}$ (assuming Poisson errors) for NGC~4762 and NGC~4754, respectively.  We estimated the fractional coverage of the GCLF by the {\it HST} data in a manner similar to our Mosaic GCLF analyses, correcting the observed GCLF for contamination by applying radially-dependent contamination corrections.  For NGC~4762, we adopted circular radial bins with a $0\farcm6$ width ranging from $0\farcm1$ to $2\farcm5$.  For NGC~4754, we adopted circular, logarithmically-spaced bins ranging from $0\farcm09$ to $2\farcm0$ to better sample the GC surface density near the galaxy center.  We fit the corrected GCLF with a Gaussian function assuming three different $g$ magnitude bin sizes (0.4, 0.5, 0.6) and fixed values for the dispersion and peak apparent magnitude.  We adopted the VCS values for the GCLF dispersions of $\sigma=1.7,1.1$ and peak apparent magnitudes of $g=24.6,24.0$ for NGC~4762 and NGC~4754, respectively \citep{vi10}.  The mean fractional coverage of the theoretical GCLF by the observed GCLF was $0.599 \pm 0.002$ for NGC~4762 and $0.859 \pm 0.001$ for NGC~4754.  Varying the choice of dispersion by $\pm 0.1$ showed a $1\%-5\%$ change in the fractional coverage, and we account for this variation in our error estimate on the total number of GCs in Section~\ref{sec_numbers}.

\section{Global Properties of the Globular Cluster System}\label{sec_global}

\subsection{Radial Distribution of the GC System}\label{sec_radial}

\begin{figure}
\plotone{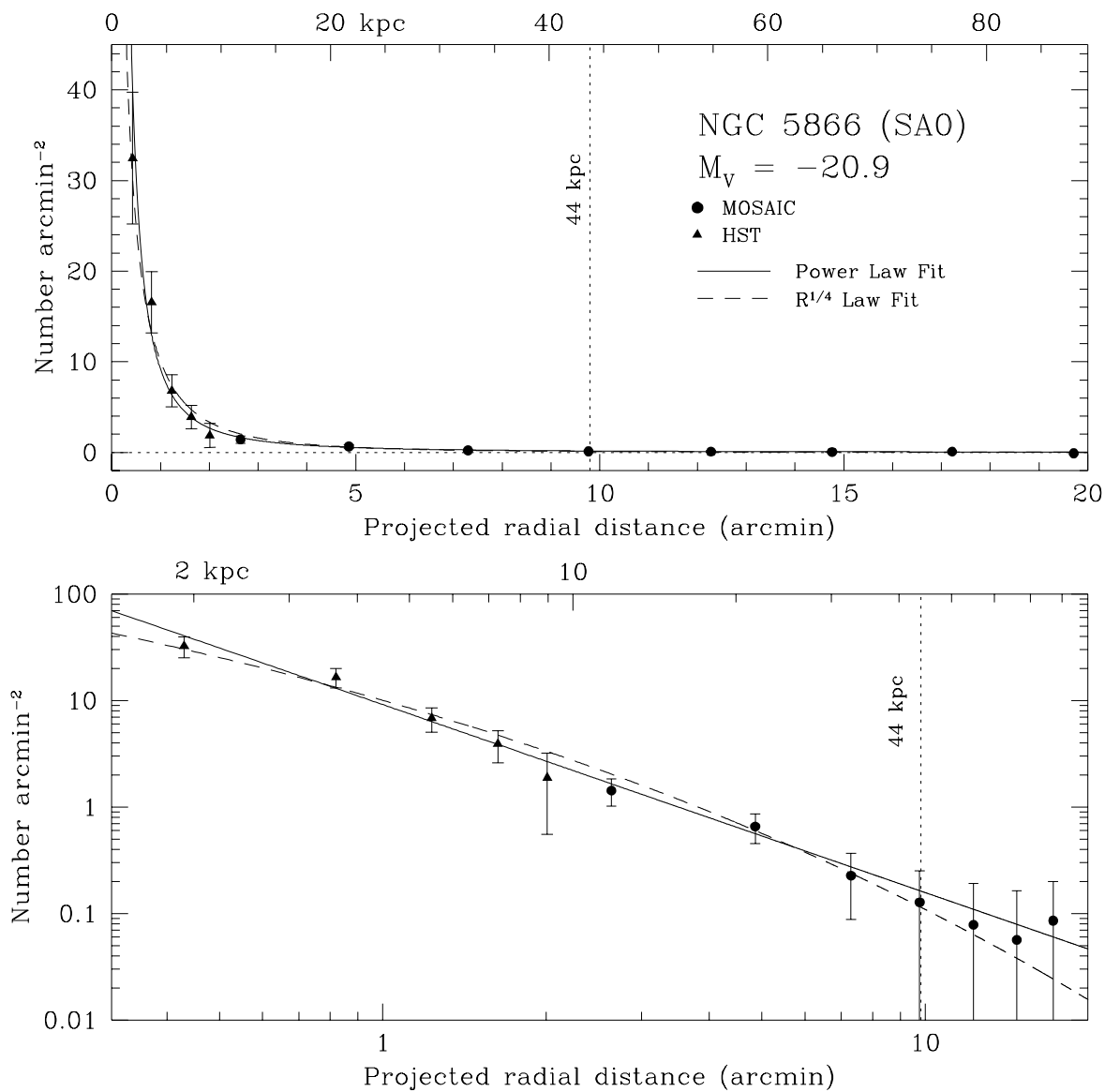}
\caption{Radial profile of the NGC~5866 GC system.  The profile has been corrected for missing area, contamination, and magnitude incompleteness (see Section~\ref{sec_radial}). The top panel shows the corrected surface density as a function of projected radial distance from the galaxy center ($r=0$). The bottom panel shows the corrected surface density as a function of projected radius on a logarithmic scale. The filled circles are the Mosaic data points from this study and the filled triangles are our analysis of the {\it HST} {ACS} GC candidates from \citet{ca07}. Both panels show the best fit power-law and $r^{1/4}$ profile fits (described in Section~\ref{sec_radial}) as the solid line and dashed line, respectively.
\label{n5866_rad_prof}}
\end{figure}

\begin{figure}
\plotone{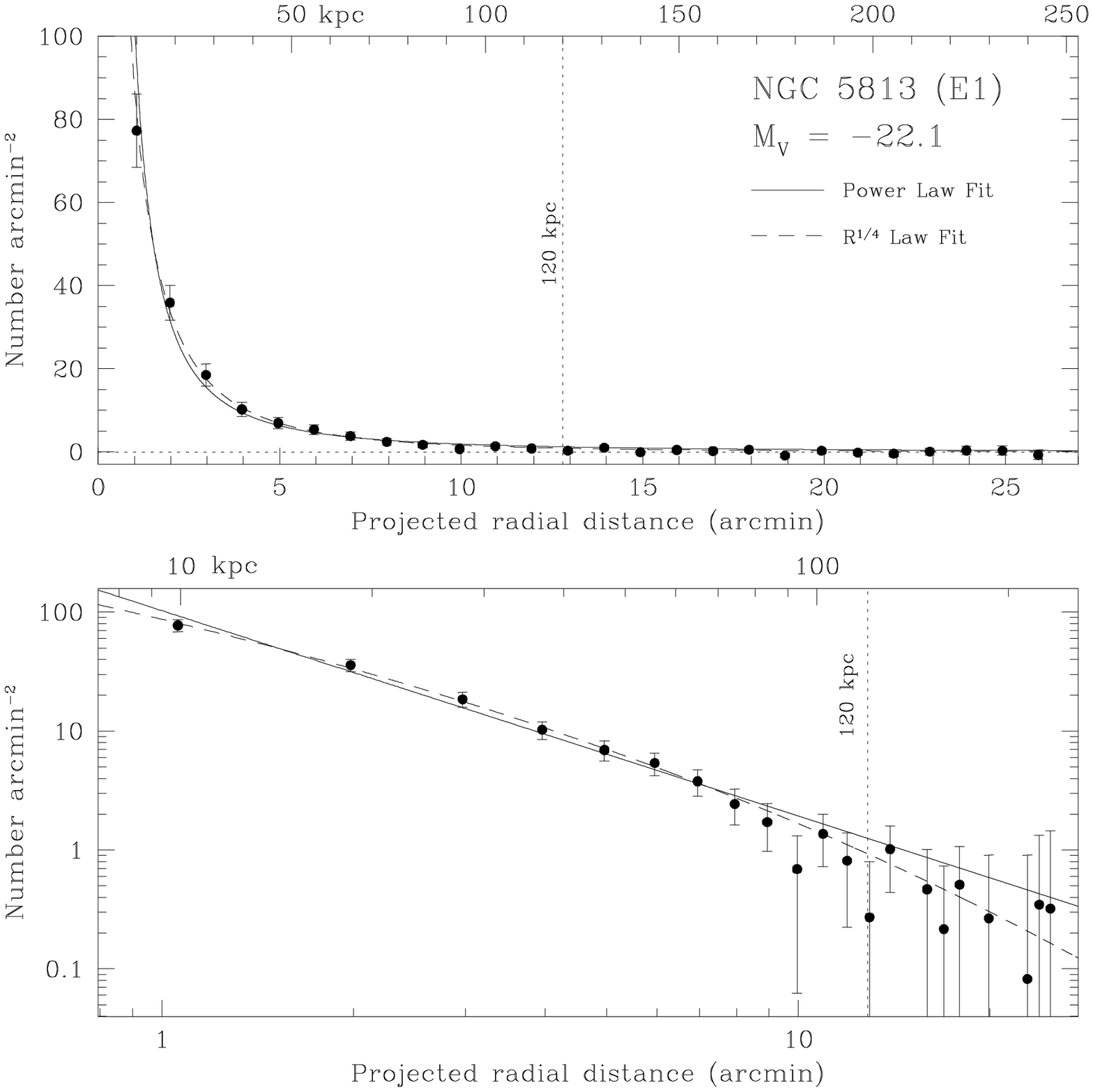}
\caption{Radial profile of the NGC~5813 GC system plotted in the same manner as Figure~\ref{n5866_rad_prof}.  See caption for Figure~\ref{n5866_rad_prof} for additional plot details. 
\label{n5813_rad_prof}}
\end{figure}

\begin{figure}
\plotone{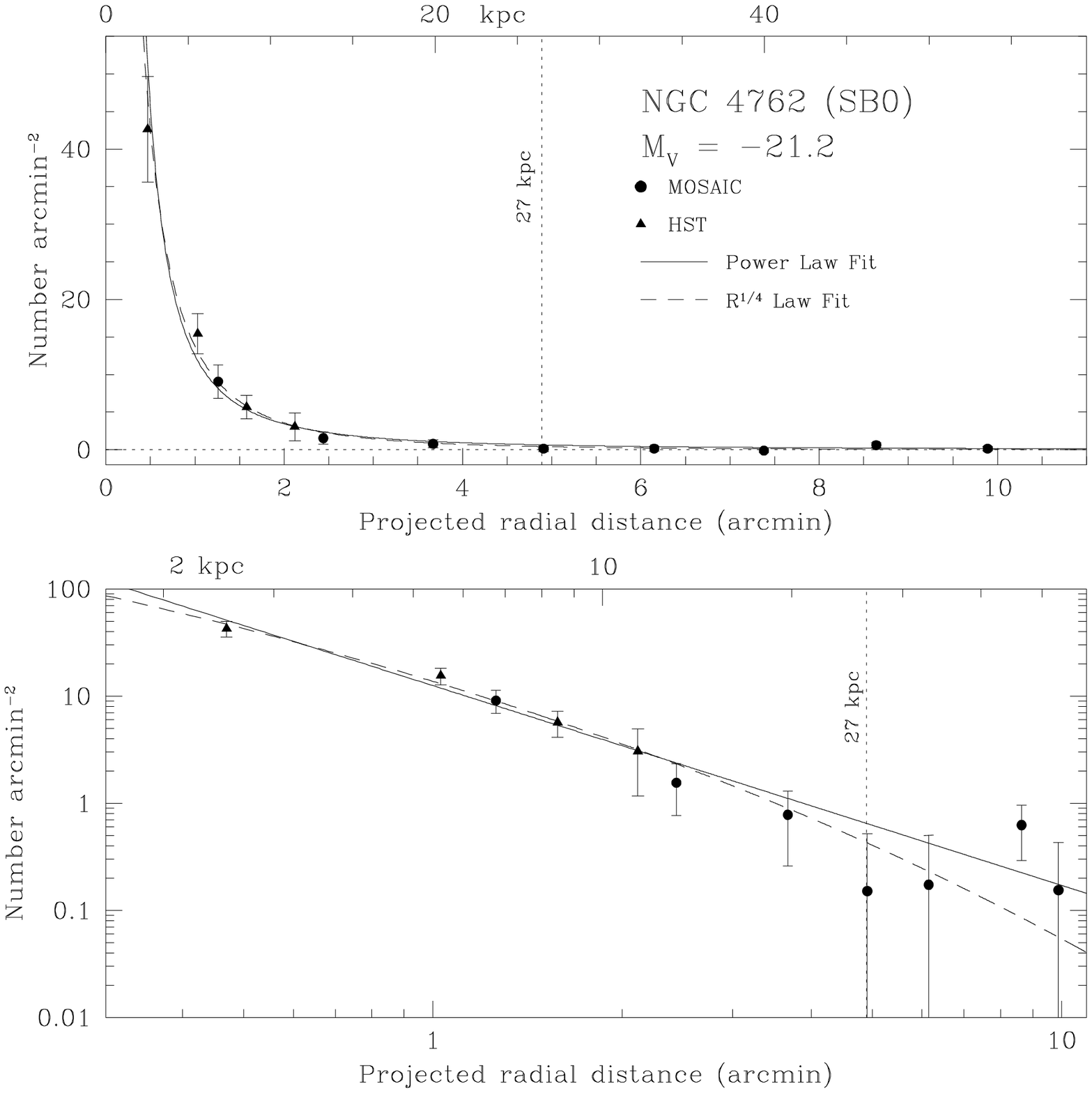}
\caption{Radial profile of the NGC~4762 GC system plotted in the same manner as Figure~\ref{n5866_rad_prof}.  See caption for Figure~\ref{n5866_rad_prof} for additional plot details.
\label{n4762_rad_prof}}
\end{figure}

\begin{figure}
\plotone{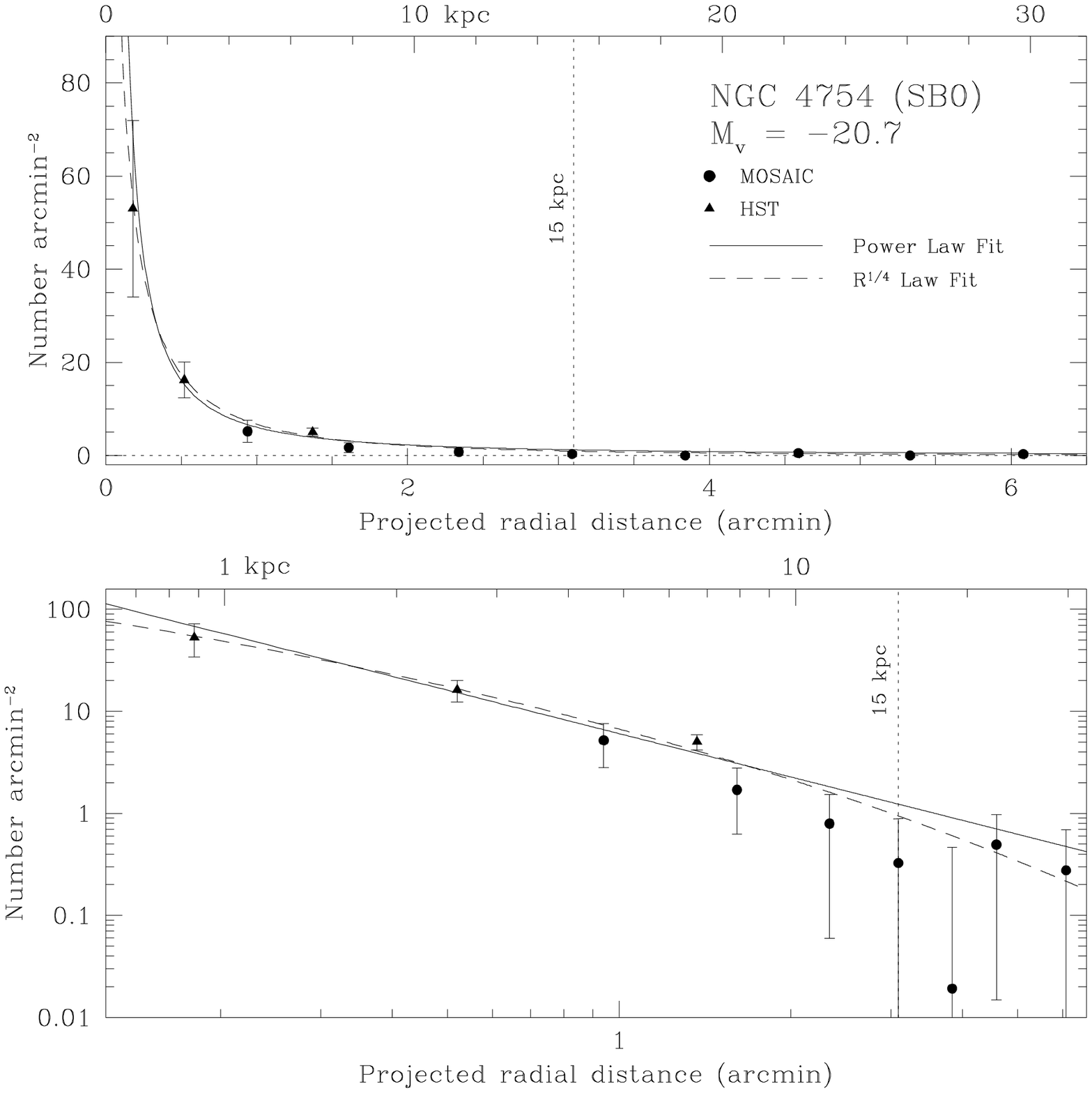}
\caption{Radial profile of the NGC~4754 GC system plotted in the same manner as Figure~\ref{n5866_rad_prof}.  See caption for Figure~\ref{n5866_rad_prof} for additional plot details.
\label{n4754_rad_prof}}
\end{figure}

\begin{figure}
\plotone{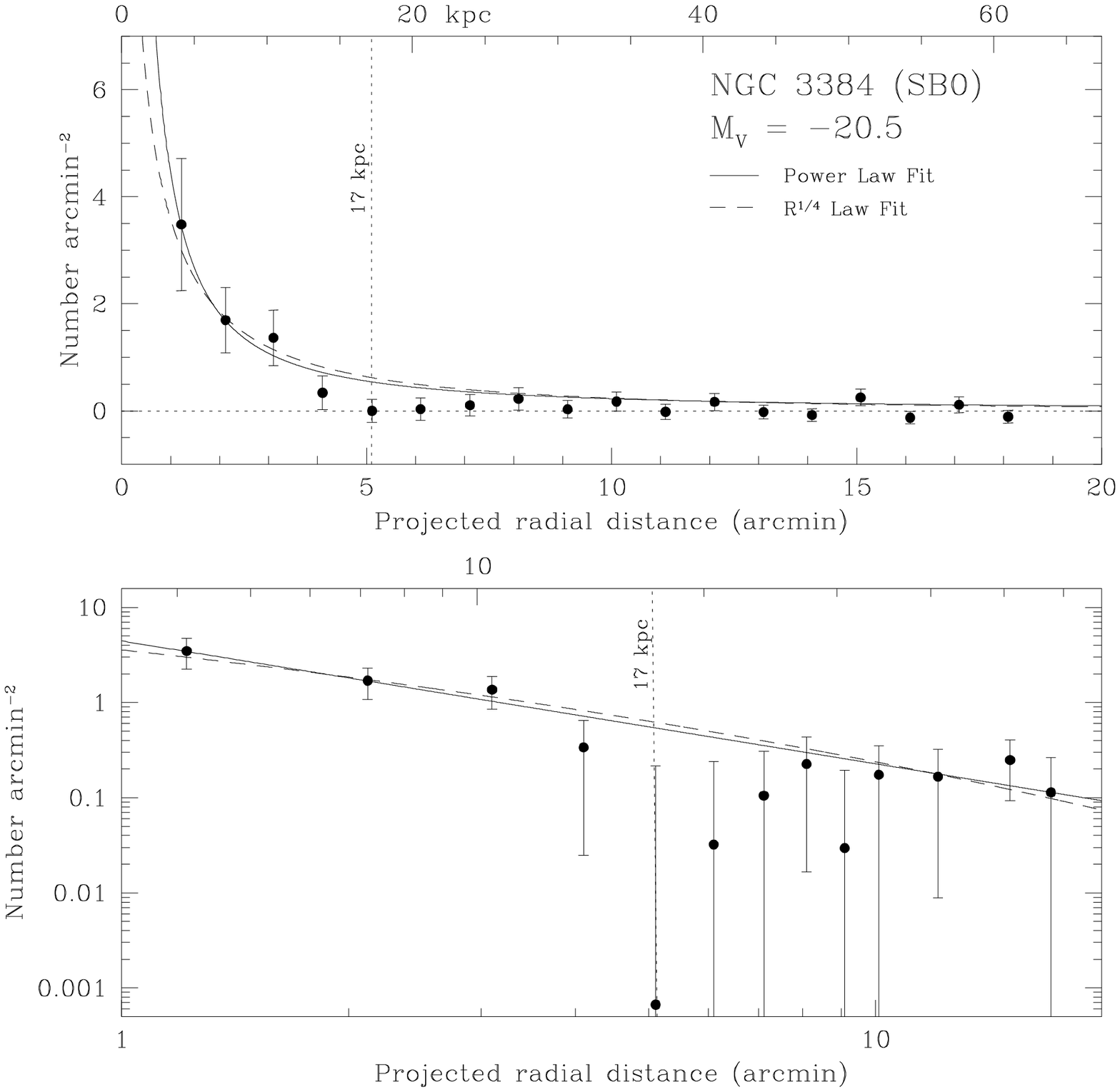}
\caption{Radial profile of the NGC~3384 GC system plotted in the same manner as Figure~\ref{n5866_rad_prof}.  See caption for Figure~\ref{n5866_rad_prof} for additional plot details.
\label{n3384_rad_prof}}
\end{figure}

The spatial distributions of the GC systems were investigated by constructing azimuthally-averaged radial surface density profiles centered on each host galaxy.  For galaxies that had additional {\it HST} data (NGC~5866, NGC~4754, NGC~4762), the {\it HST} and Mosaic radial profiles were constructed independently.   For each GC system, we binned the GC candidates in concentric circular annuli. For both the Mosaic and {\it HST} data, we adopted the same radial bins chosen when we used the initial radial profiles to determine the contamination level (see Sections~\ref{sec_asymp} and \ref{sec_hst}).   The final radial profiles were corrected for missing area, contamination, and magnitude incompleteness as follows.  For each circular annulus we accounted for the missing area by computing the effective area of the radial bin.
We correct the profile for contamination by adopting the radially dependent contamination fractions, applying the values determined in Section~\ref{sec_asymp} for the Mosaic data and in Section~\ref{sec_hst} for the {\it HST} data.  Finally, we corrected the radial profiles for magnitude incompleteness by dividing each radial bin by the fractional coverage of the observed Mosaic or {\it HST} GCLF (see Sections~\ref{sec_gclf} and~\ref{sec_hst}, respectively).  The surface density was then computed using the effective area and the corrected total number of GCs in each bin.  The uncertainties in the surface density are calculated assuming Poisson errors in the number of GCs and contaminating objects.  In Tables~\ref{tbl_n5866_prof} to~\ref{tbl_n3384_prof} we list the final, corrected radial surface density data for the six galaxies in this study: mean radius of the bin, corrected surface density of GCs and uncertainty, fractional area coverage of the annulus, and the data source (either {\it HST} or Mosaic). Figures~\ref{n5866_rad_prof} to~\ref{n3384_rad_prof} show the radial surface density of GC candidates as a function of the projected radial distance.

The radial surface density profiles follow the trends observed in our previous wide-field imaging studies: a strong overdensity of GCs centered on the host galaxy which decreases smoothly as a function of projected radius away from the galaxy.  As noted in Section~\ref{sec_asymp}, the profiles decrease to a constant surface density (consistent with zero after applying the corrections), which indicates that we have observed the full radial extent of the galaxy's GC system.  For the three galaxies with both {\it HST} and Mosaic data (NGC~5866, NGC~4762, NGC~4754) we find excellent agreement between the radial profiles in the regions of overlapping spatial coverage.  In all three cases, the radial profiles continue to decrease outside the {\it HST} field of view as evidenced by the Mosiac data. To further describe the radial distribution of GCs, the corrected radial profiles were fit with both a power-law profile ($\log{\sigma_{\rm GC}}=a_0 + a_1\log{r}$) and a de Vaucouleurs profile ($\log{\sigma_{\rm GC}}=a_0 + a_1 r^{1/4}$).  The resulting fit parameters are listed in Table~\ref{tbl_prof_fits}. The integration of the best-fit radial profiles to estimate the total number of GCs is described in Section~\ref{sec_numbers}.

We estimate the radial extent of the galaxy's GC system by examining the point in the final corrected radial profile where the surface density becomes consistent with zero.  For our survey, we have defined the radial extent of the GC system as the radius at which the surface density of GCs falls to zero within the errors and stays consistent with zero (RZ07).  Although this  definition is sensitive to the survey methods (corrections for magnitude incompleteness and contamination in particular), it allows for a systematic and internally consistent comparison of the projected size of the GC system for galaxies in our survey.  For the galaxies in this study, the large spatial coverage of the Mosaic observations resulted in stochastic variations in the surface density around zero in the outer bins of the radial profile.  For example, the profile of NGC~3384 shows a small  variation above the background in a radial bin more than $12\arcmin$ from the main spatial overdensity of the GC system.  Accounting for the observed variations, we adopt the following radial bins as the radial extent of the GC systems: $r=9\farcm8$ ($44 \pm 11$ kpc) for NGC~5866, $r=12\farcm9$ ($120 \pm 14$ kpc) for NGC~5813, $r=4\farcm9$ ($27 \pm 7$ kpc) for NGC~4762, $r=3\farcm1$ ($15 \pm 4$) for NGC~4754, and $r=5\farcm1$ ($17 \pm 4$ kpc) for NGC~3384.  Our uncertainty in the radial extent includes the uncertainty in the distance modulus and a one bin-width uncertainty in our determination of where the surface density of GCs falls to zero.

\subsection{Color Distributions and Color Gradients}\label{sec_color}

An important goal of our wide-field imaging survey is to quantify the global color distribution of the GC systems of giant galaxies.  For stellar populations with ${\rm ages} \gtrsim 2~{\rm Gyr}$, broadband optical colors trace metallicity \citep{wo94,br03} such that old, metal-poor populations have photometrically bluer optical colors than old, metal-rich populations.  The presence of broad and/or bimodal color distributions of GC systems in giant galaxies -- including the Milky Way \citep{zi85} and many giant ellipticals \citep{ku01a,ku07,st07} -- 
 has therefore been largely interpreted as tracing an underlying bimodal {\it metallicity} distribution in GC populations (see Brodie \& Strader 2006 and references therein). 
Because metallicity differences in the subpopulations points to different episodes of star formation, the bimodal color distributions of GC systems have been a key differentiator between galaxy formation scenarios \citep{as92,fo97,co98,be02,mu10}.
Quantifying the properties of GC system color distributions for a large number of galaxies over a range of mass, morphology, and environment provides a key means of testing galaxy formation scenarios.

To investigate the color distributions of the galaxies in our sample, we 
construct and analyze a sample of GCs that have 90\% magnitude completeness in all three filters.  We minimize the effects of contamination by performing a radial cut on the 90\% complete sample, limiting our analysis to only those GCs that are within the radial extent of the GC system. For each galaxy, we analyze the resulting sample
for color bimodality using the KMM mixture modeling code \citep{as94} to test whether the distribution is better described statistically with one more Gaussian distributions.  We explore both homoscedastic (same dispersion) and heteroscedastic (differing dispersion) models. We plot the $B-R$ colors for the radially-cut, 90\% GC sample as a function of projected radius from the galaxy center to explore the possibility of color gradients in the galaxy's GC. Linear least-squares fits to these data are used to evaluate the presence of color gradients.

\begin{figure}
\plotone{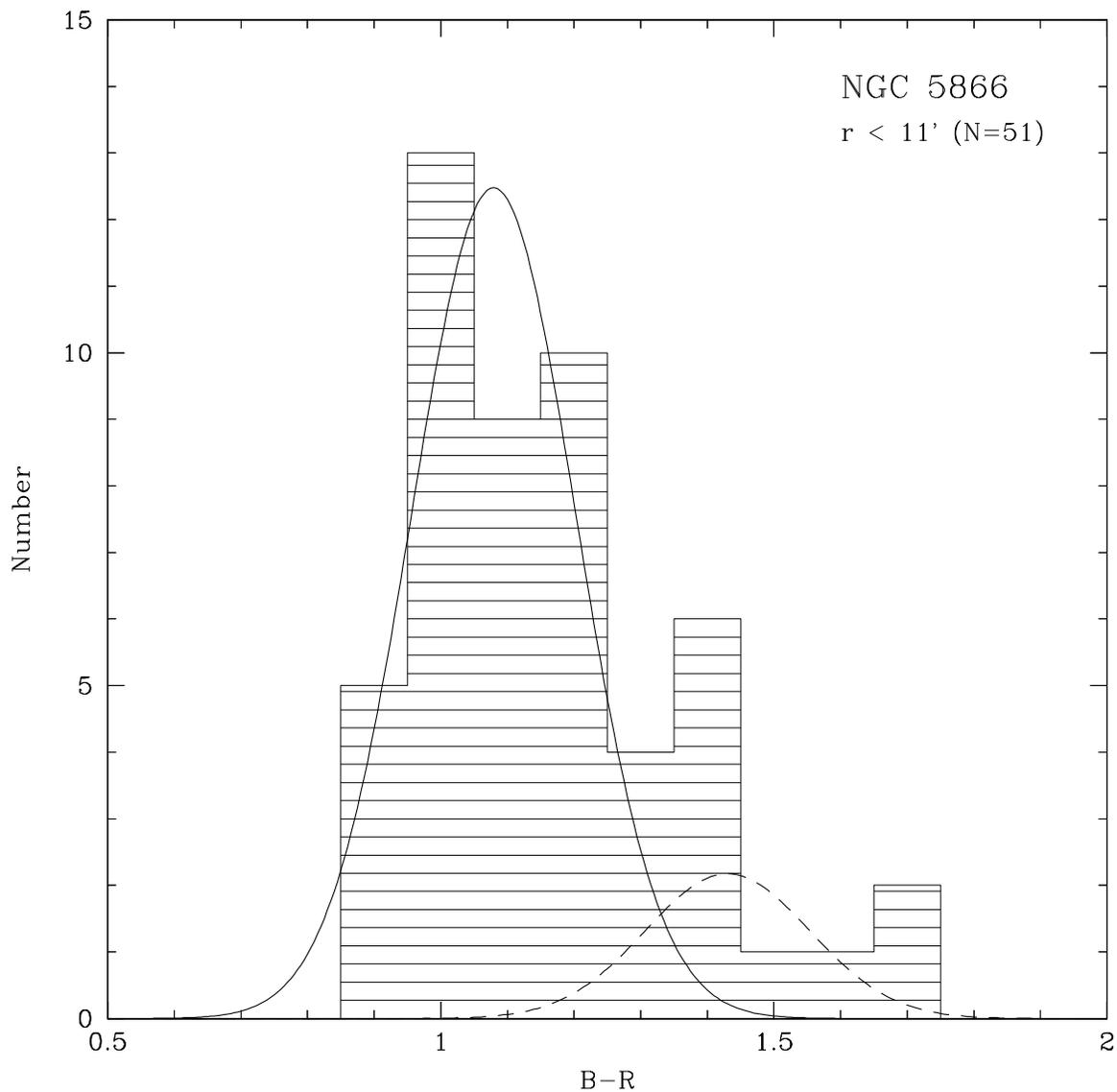}
\caption{Color Distribution and KMM mixture modeling results for NGC~5866.  The shaded histogram shows the $B-R$ color distribution for the radial cut ($r<11\arcmin$), 90\% magnitude complete sub-sample of 51 Mosaic GC candidates for NGC~5866 (see Section~\ref{sec_color}).   The KMM mixture modeling results are shown as the solid and dashed curves.  The solid curve denotes the blue (metal-poor) population and the dashed curve denotes the red (metal-poor) population.
\label{n5866_color_dist}}
\end{figure}

\begin{figure}
\plotone{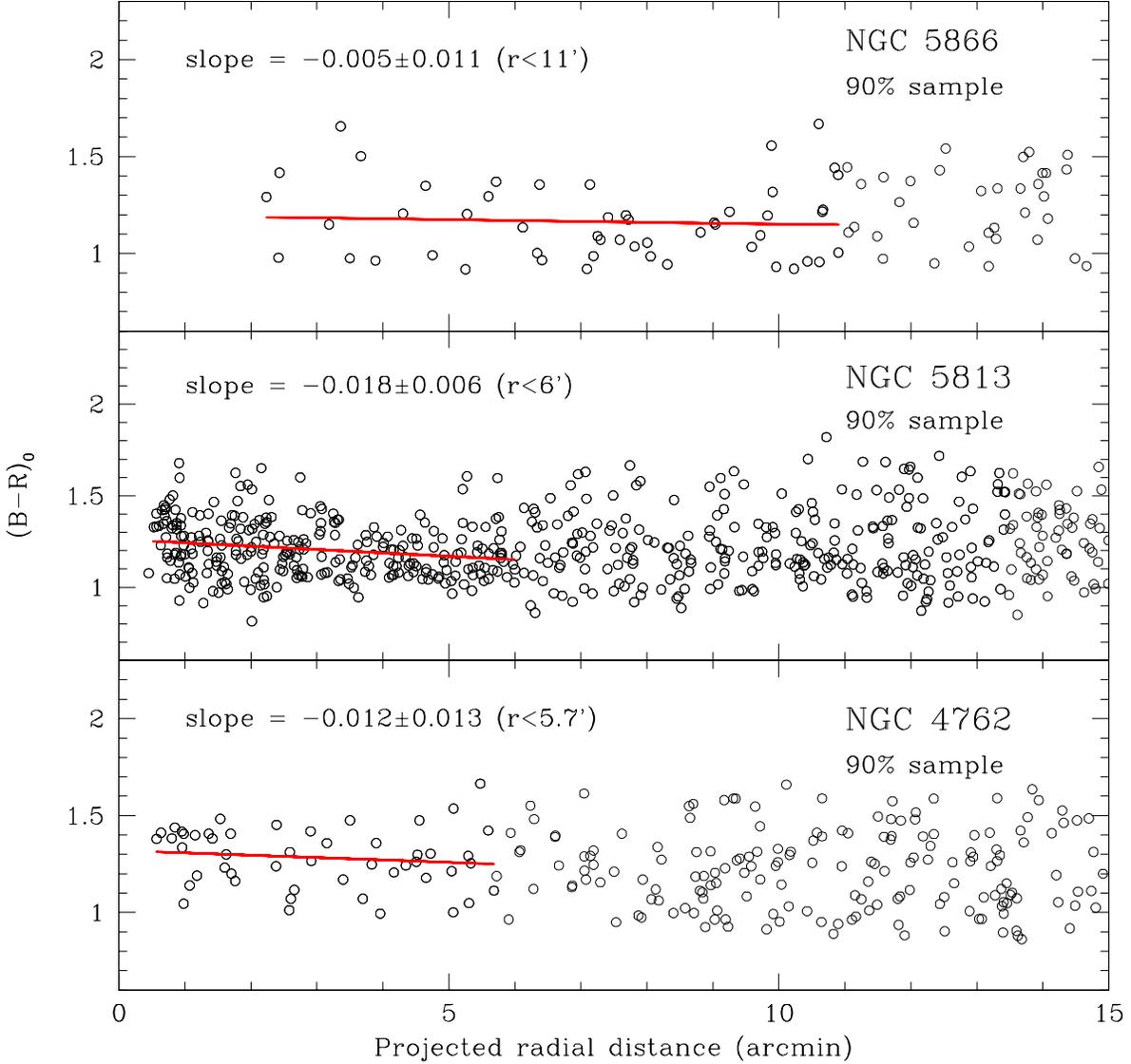}
\caption{$B-R$ color versus projected radial distance of the 90\% sample of GC candidates for NGC~5866, NGC~5813, and NGC~4762.  GC candidates in the 90\% sample are shown as open circles.  For NGC~5866 and NGC~4762, we performed linear fits to the data using only the GC candidates within the radial extent of the GC system and find no evidence of statistically signficant color gradients.  For NGC~5813, we performed linear fits over the full radial extent  ($r<13\farcm4$) and in the inner regions of the GC system.  We find evidence of a statistically significant color gradient in the inner $6\arcmin$ ($\sim 55$ kpc) of the GC system (slope = $\Delta (B-R)/\Delta r$ = $-0.018\pm0.006$).  For each galaxy we show the linear fit as a red solid line and the best-fit slope and $1\sigma$ uncertainty.   NGC~4754 and NGC~3384 had insufficient numbers of GC candidates in the 90\% sample (within the radial extent of the GC system) so no color gradient analysis was performed.
\label{group_color_grad}}
\end{figure}

We discuss the results for individual galaxies here (The results regarding color bimodality and color gradients are summarized in Table~\ref{tbl_ngc}): 

{\it NGC~5866}. -- For NGC~5866, the 90\% complete, radially-cut ($r<11\arcmin$) sample resulted in a total of 51 objects, just above the 50 objects necessary for reliable KMM results.  The color distribution, shown in Figure~\ref{n5866_color_dist}, is quite broad and has peaks at $B-R\sim 1.0 $ and $B-R~\sim 1.4$.  The KMM results indicate that the distribution is likely bimodal at the $97\%$ ($>2\sigma$) confidence level (homoscedastic case), finding a blue (lower metallicity) peak at $B-R=1.08$ and a red peak (higher metallicity) at $B-R=1.43$.  The KMM analysis finds a split in the GC colors at $B-R\sim 1.3$, with $76\%~ (N=39)$ GCs comprising the blue subpopulation and $24\%~(N=12)$ GCs comprising the red subpopulation.  Figure~\ref{n5866_color_dist} shows the KMM results with the 90\% complete, radial cut GC color distribution.  Separating the GC sample at the empirically-determined color separation between blue and red subpopulations for elliptical galaxies of $B-R=1.23$ (RZ01, RZ04, Kundu \& Whitmore 2001a), we find similar results: a blue fraction of $71\%~ (N=36)$ and a red fraction of $29\%~ (N=15)$.  CBR07 also performed a KMM analysis of their $B-R$ distribution of 109 GC candidates.  They find a likelihood that the distribution is bimodal at the $96\%$ confidence level, with distribution peaks at $B-R=1.12$ and $B-R=1.35$ and an equal number of red and blue GCs in each subpopulation.  For our analysis of the mass-normalized number of blue GCs ($T_{\rm blue}$), we average the KMM and empirical split results and adopt a blue GC fraction of $74\%$.  An examination of the $B-R$ colors for the radial cut, 90\% sample as a function of projected radius (see Figure~\ref{group_color_grad}) showed no statistically significant color gradient within $r<11\arcmin$ (the radial extent of the GC system).  CBR07 find a $B-R$ color gradient in their sample of {\it HST} GC candidates within $r < 2\arcmin$ resulting from the higher central concentration of the red GC population in the inner $\sim 1\arcmin$.

\begin{figure}
\plottwo{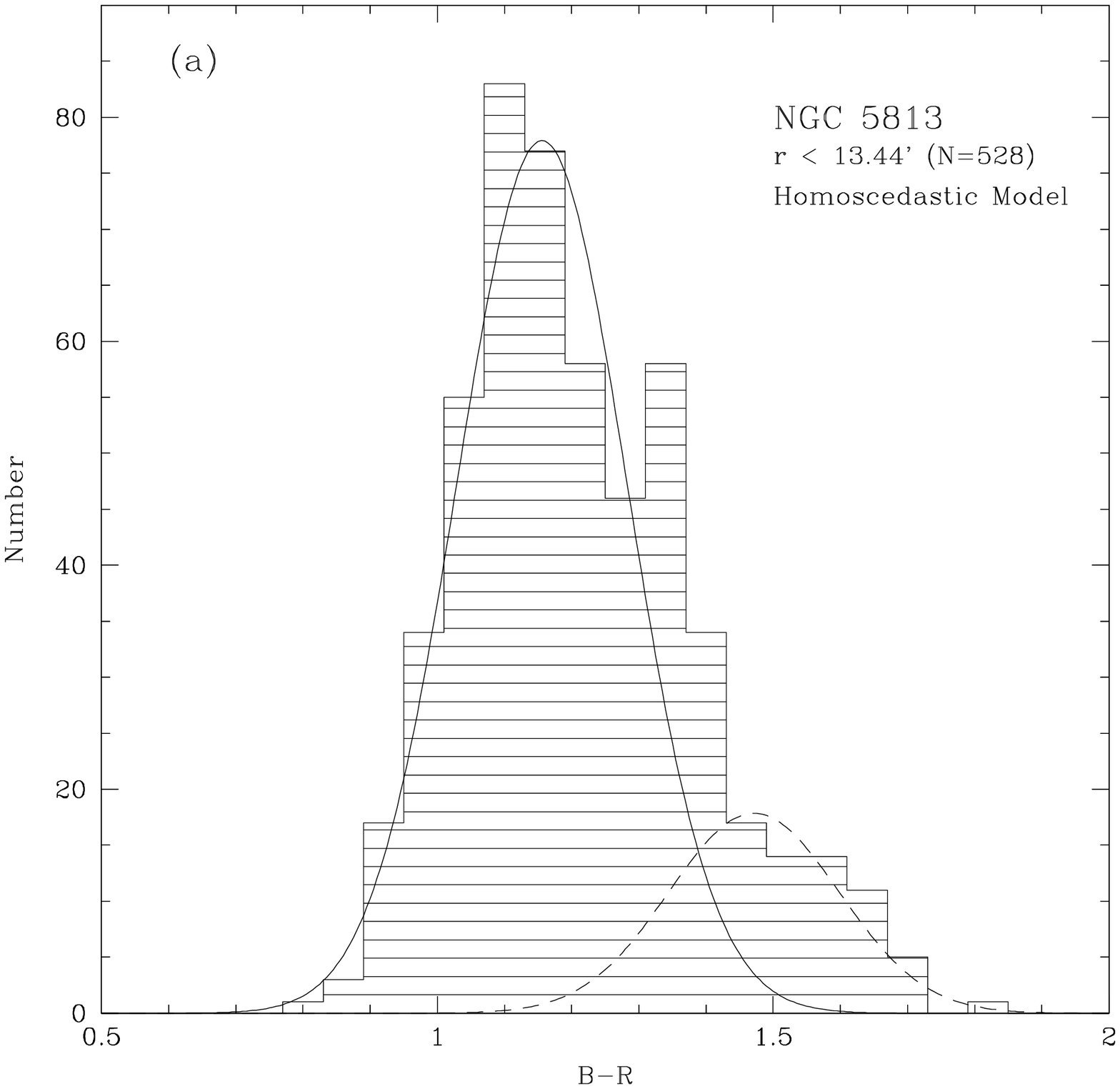}{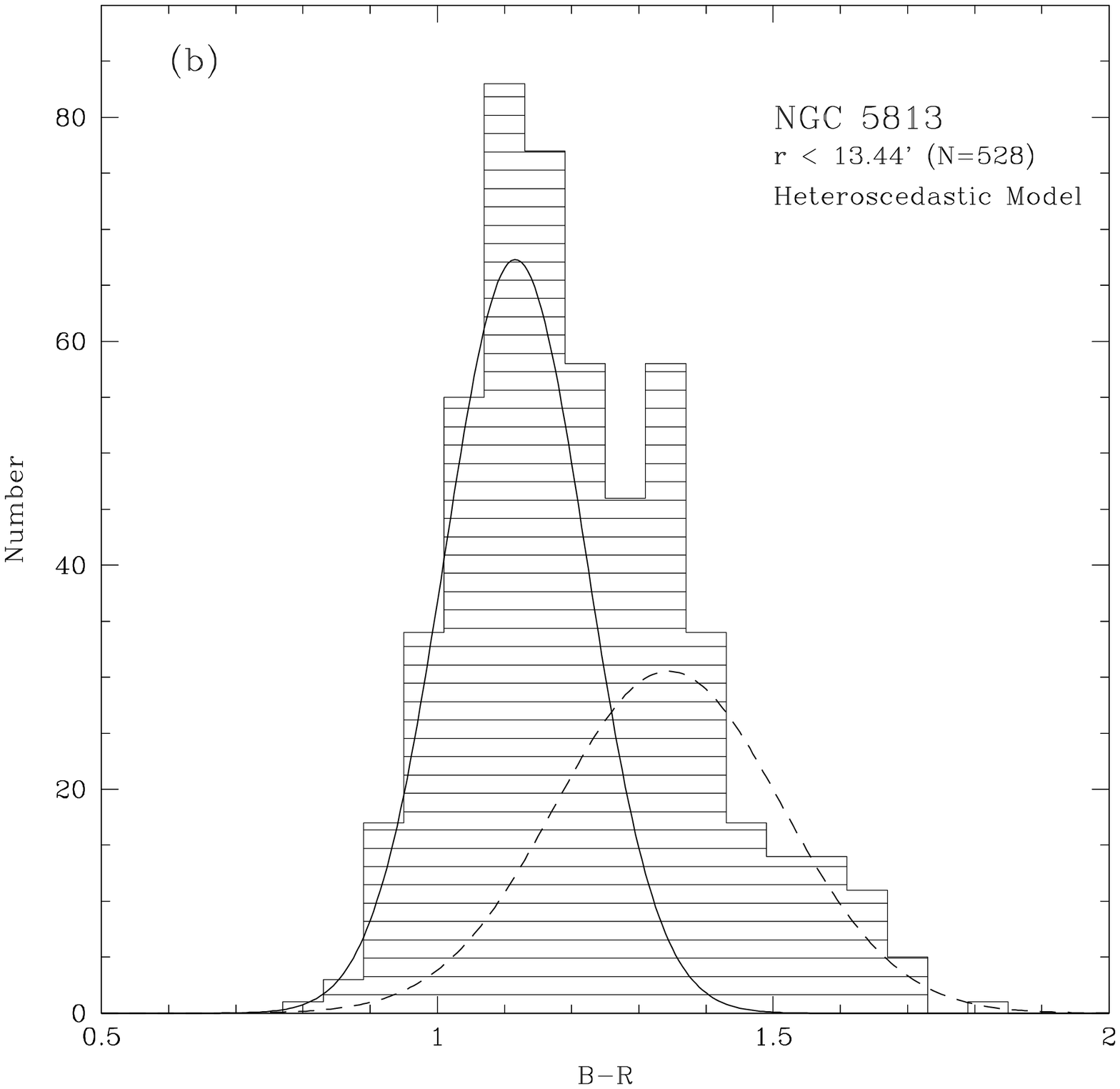}
\caption{Color Distribution and KMM mixture modeling results for NGC~5813.  The shaded histograms (both panels) show the $B-R$ color distribution for the radial cut ($r<13\farcm44$), 90\% magnitude complete sub-sample of 528 Mosaic GC candidates (see Section~\ref{sec_color}).  The KMM mixture modeling results for the homoscedastic (same dispersions) case in panel (a) and the results for the heteroscedastic case (differing dispersions) case in panel (b).  In both panels the solid curves denote the blue (metal-poor) population and the dashed curves denote the red (metal-poor) population.
\label{n5813_color_dist}}
\end{figure}

{\it NGC~5813}. --  The 90\% complete, radial cut sample ($r<13\farcm4$) for NGC~5813 contained 528 objects. Figure~\ref{n5813_color_dist} shows the $B-R$ color distribution for this sample; the distribution is broad and has peaks at $B-R\sim 1.1$ and $B-R\sim 1.35$, including a red ``tail" of objects with $B-R>1.5$.  The KMM results indicate that unimodality can be ruled out at better than the $99.99\%$ confidence level in both the homoscedastic and heteroscedastic cases.  For both mixture models, the blue peak is found at roughly the same location ($B-R\sim1.1-1.2$), while the location of the red peak differs depending on the dispersion model assumptions ($B-R\sim1.5$ for the homoscedastic case versus $B-R\sim1.3$ for the heteroscedastic case).  The resulting number of red and blue GCs comprising each subpopulation also changes; we find a blue fraction of $83\%$ ($N=440$) in the homoscedastic case compared to $63\%$ ($N=331$) in the heteroscedastic case.  Splitting the distribution at the empirically-determined color separation of $B-R=1.23$ we find a blue fraction of $58\%$ ($N=304$), in good agreement with the heteroscedastic KMM results.   We adopt the mean of the three blue fraction values (two KMM models and the empirical split) of $68\%$ as our final estimate of the blue fraction.  

\begin{figure}
\plotone{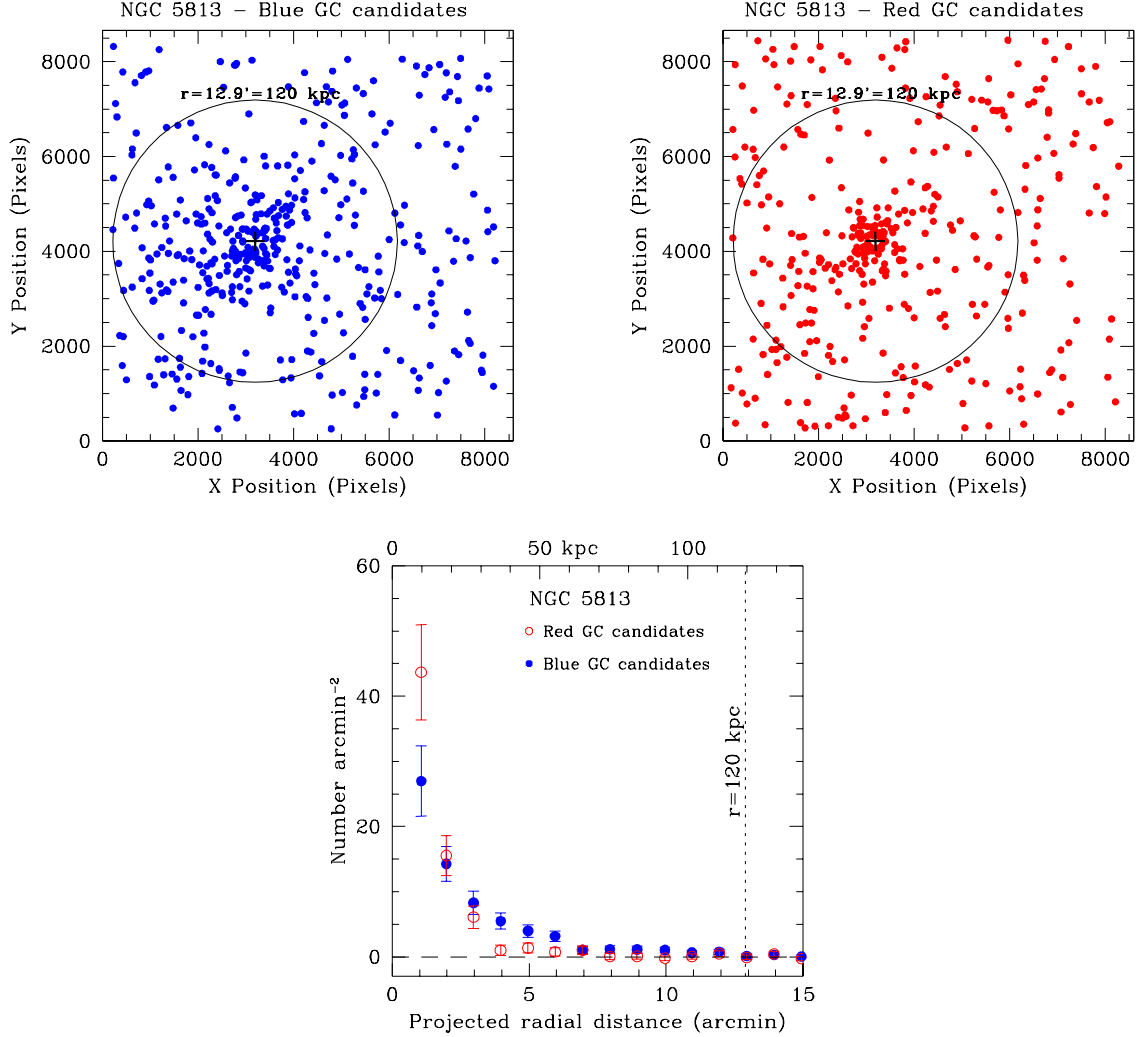}
\caption{Spatial distribution of the red and blue GC subpopulations in NGC~5813.  The upper panels shows the spatial positions of the red (metal-rich) and blue (metal-poor) GC candidates on the Mosaic frame as blue filled circles (upper left panel) and red filled circles (upper right panel).  The center of NGC~5813 is shown as the black cross. The radial extent of the full GC system (red+blue subpopulations; see Figure~\ref{n5813_rad_prof}) is shown as the circle of radius $r=12\farcm9$ ($120$ kpc).  The lower panel shows the radial surface density profiles of the two subpopulations.  We show only the inner $15\arcmin$ of the profile; the complete radial profile for the full (blue+red) GC population is shown in Figure~\ref{n5813_rad_prof}.  The red GC subpopulation is shown as the red open circles and the blue subpopulation is shown as the blue filled circles.  The spatial distributions of the subpopulations are discussed in Section~\ref{sec_color}.
\label{n5813_spatial_color}}
\end{figure}

With a large number of GC candidates in NGC~5813 and clear evidence for bimodality, we can extend this analysis to look for differences in the spatial distributions of the red and blue GC subpopulations.  We used the empirically-determined color separation of $B-R=1.23$ to split the GC candidates in the full 90\% sample  (no radial cut; $N=809$ objects) into red and blue subpopulations. Figure~\ref{n5813_spatial_color} shows the spatial positions of the GCs for each subpopulation with respect to the host galaxy center.  The red subpopulation is noticably more centrally concentrated compared to the blue subpopulation.  This is consistent with similar studies of the spatial distributions of GC subpopulations in other giant galaxies (see Brodie \& Strader 2006 and references therein).   We constructed radial surface density profiles (corrected for contamination, magnitude incompleteness, and missing area) for the two subpopulations (bottom panel of Figure~\ref{n5813_spatial_color}).  The radial profile for the red GC subpopulation is steeper than that for the blue GC subpopulation and becomes consistent with zero (within the errors) at $\sim 8\arcmin$ ($\sim 75$ kpc).  The blue GC subpopulation, however, is more spatially extended and the profile does not becomes consistent with zero until $\sim 13\arcmin$ ($\sim 120$ kpc) -- the radial extent of the {\it combined} GC profile as determined in Section~\ref{sec_radial}.  An examination of the $B-R$ colors as a function of projected radius (see Figure~\ref{group_color_grad}) shows a statistically significant color gradient (slope = $\Delta (B-R)/\Delta r = -0.018 \pm 0.006$) in the inner $6\arcmin$ ($\sim 55$ kpc) of the GC system.  This results from the decreasing fraction of red GCs as a function of radius from the host galaxy center within this region of the GC system.  Using the GC $B-R$ color-[Fe/H] relation from Rhode \& Zepf (2001, 2004; derived from the Milky Way GC data of Harris 1996), we translated the GC colors to [Fe/H] and fit the metallicities as function of radius.  This yields a best-fit metallicity gradient of $\Delta {\rm [Fe/H]}/\Delta {\log r} = -0.36 \pm 0.10$. If we consider the entire radial extent of the GC system ($r<13\farcm4$), we find no evidence of a statistically significant color gradient.

{\it NGC~4762}. -- For NGC~4762, we constructed a 90\% complete, radial cut sample of $N=50$ objects.  Because the radial extent of the GC system (outer edge at $r=5\farcm5$) included fewer than 50 GC candidates (only $N=44$), we extended the radial cut to $r=5\farcm7$ to create a sample of exactly $N=50$ objects.  The $B-R$ color distribution is quite broad and shows peaks at $B-R\sim 1.25$ and $B-R\sim1.4$.  The radial distribution of these GCs (see Figure~\ref{group_color_grad}) shows that the $B-R\sim1.4$ peak is due to a cluster of 10 GC candidates at projected radii $r<2 \arcmin$ with $1.35<B-R<1.45$.  These objects are likely a population of ``faint fuzzies" (FFs): star clusters that are typically old, metal-rich, have slightly larger sizes than typical GCs, and are often spatially associated with the host galaxy disk (Brodie \& Larsen 2002; Peng et al. 2006b and references therein).  The KMM analysis of the $N=50$ sample shows that we cannot rule out unimodality at a statistically significant level; the distribution is only consistent with bimodality at the $\sim 54\%$ significance level (mean significance level of the heteroscedastic and homoscedastic models).  The ACSVCS study of the $g-z$ color distribution of NGC~4762 \citep{pe06} also did not detect bimodality at a statistically significant level in their sample of $N=123$ GC candidates.  Given the lack of strong evidence for color bimodality, we estimate the blue fraction of GCs using the empirically determined split at $B-R=1.25$.  For the $N=50$ sample (with radial cut at $r=5\farcm7$) we find a fraction of blue GCs of $34\%$ ($N=17$).  If we remove the ten FFs from the sample, we find a blue GC fraction of $43\%$ ($N=17$) using the empirically-determined split.  We therefore adopt a blue GC fraction of $\sim 40\%$.  We do not require bimodality to estimate a meaningful blue fraction; simply using the empirical split allows us to compare the blue GC population in a self-consistent manner. Examining the $B-R$ colors of our 90\% complete, radial cut sample as a function of projected radius (see Figure~\ref{group_color_grad}), we find no evidence of a statistically significant color gradient within $r=5\farcm7$ in NGC~4762 (either with or without the FF population).  The ACSVCS study of $g-z$ color gradients (limited to the inner $2\farcm4$) also did not find evidence of a color gradient in NGC~4762 \citep{li11}.

{\it NGC~4754}. -- For NGC~4754, the 90\% sample based on a radial extent of the GC system ($r=3\farcm5$) contained only $N=24$ objects.  To create a sample of the minimum $N=50$ objects for KMM analysis we would have to extend the radial cut to $r=6\farcm1$, thereby including a large number of potentially contaminating objects in the sample. We therefore apply the empirically-determined cut at $B-R=1.23$ and find a blue fraction of $33\%$ ($N=8$).  Increasing the radial cut by 1-2 radial bins (an additional $0\farcm75-1\farcm5$) does not significantly change this result even though the sample increases to $N=39$ objects. The ACSVCS study NGC~4754's $g-z$ color distribution \citep{pe06} only detected bimodality at less than the $2\sigma$ significance level ($89\%$) using $N=83$ GC candidates. They estimate a slightly larger blue GC fraction of $50\%$.  Although we find no strong statistical evidence for color bimodality, we will adopt a blue GC fraction of $\sim 40\%$ (average of the empirical split and ACSVCS results) in our discussion of the mass-normalized number of blue GCs ($T_{\rm blue}$) in Section~\ref{sec_numbers}. Examining the $B-R$ colors for the 90\% sample with various radial cuts, the lack of sufficient numbers of GC candidates precludes definitive conclusions about a color gradient in the NGC~4754 GC system.  In addition, we find no evidence for a FF population similar to that in NGC~4762.

{\it NGC~3384}. -- For NGC~3384, the number of GC candidates in the 90\% complete, radial cut sample was insufficient to reliably run the KMM tests ($N=13$).  With no radial cut the 90\% complete sample only consisted of $N=45$ objects, which is still below the recommended number of objects ($N=50$).  We therefore used the empirically-determined split at $B-R=1.23$ to estimate the blue GC fraction.  With no radial cut, the 90\% complete sample ($N=45$) has 30 objects with $B-R<1.23$, yielding a blue GC fraction of 67\%.  Relaxing the radial cut imposed at the radial extent of the GC system ($r=5\farcm6$; outer edge of bin which defines the radial extent) by an additional two radial bins (to $r=7\farcm6$) still yields only $N=18$ objects, of which 55\% ($N=10$) are bluer than $B-R=1.23$.  As our final blue GC fraction estimate for NGC~3384, we adopt the average of the estimates from the full 90\% sample (with no radial cut) and the 90\% sample with a radial cut at $r=7\farcm6$, finding a value of $\sim60\%$.  Examining the $B-R$ colors for the 90\% sample with various radial cuts, the lack of sufficient numbers of GC candidates precludes definitive conclusions about a color gradient in the NGC~3384 GC system.

\subsection{Total Numbers of GCs and Global Specific Frequencies}\label{sec_numbers}

To estimate the total number of GCs ($N_{\rm GC}$) for each galaxy, we integrate the best-fit radial surface density profiles as determined from the profile fitting in Section~\ref{sec_radial}.  For each galaxy, we integrate the profile from the inner edge of the innermost bin to the outer edge of the bin which defines the radial extent (the bin where the surface density first becomes consistent zero within the errors and stays consistent with zero).  Because we have no information regarding the surface density inwards of the innermost radial bin (due to the high surface brightness of the galaxy), we estimate the inner region contribution to $N_{\rm GC}$ as follows.  We consider two possible scenarios: that the best-fit profile continues inward to the galaxy center or that the profile could remain flat (constant surface density) in the inner region.
We then take the average of these two estimates as the number of GCs contributed by the innermost, unmeasured region. The uncertainty on $N_{\rm GC}$  
includes the following sources of error: (1) Poisson errors on the number of GCs and contaminating objects; (2) fractional coverage of the GCLF, based on variations due to choices in bin size, dispersions, and bin centers; (3) the number of GCs assumed for the central region of the galaxy.  For NGC~5866 and NGC~3384, we include an additional error term due to the larger uncertainty in the radial extent of the GC system in these galaxies. 

We calculate the luminosity- and mass-normalized specific frequencies using our estimates of $N_{\rm GC}$, mass-to-light ratios $(M/L)_V$ from \citet{ze93}, and galaxy absolute magnitudes from the RC3 (see Table~\ref{tbl_targets}).  The luminosity-normalized specific frequency $S_N$ is defined as the total number of GCs normalized by the $V$ band luminosity of the galaxy \citep{ha81}, or,

\begin{equation}
S_N \equiv N_{\rm GC}10^{0.4(M_V+15)}.
\end{equation}

\noindent Normalizing the total number of GCs by the stellar mass of the galaxy $M_{\rm G}$ defines the mass-normalized specific frequency $T$ \citep{ze93}:

\begin{equation}
T \equiv \frac{N_{\rm GC}}{M_{\rm G}/10^9 M_\odot}.
\end{equation}

\noindent When calculating values of $T$, we assume the mass-to-light ratios $(M/L)_V$ of $7.6$ for lenticulars 
and $10.0$ for ellipticals \citep{ze93}.  In addition to the uncertainty in $N_{\rm GC}$, the errors in $S_N$ and $T$ include an uncertainty in the galaxy magnitude.  We adopted a galaxy magnitude uncertainty of $3$ times the standard error in the RC3 magnitude (typically $\sim 0.2$ mag; \citealt{RC3}) and the uncertainties in all contributing quantities were added in quadrature to give the total error in $S_N$ and $T$.  For all galaxies, we find that the uncertainty in the galaxy magnitude is the dominant source of error in the specific frequencies.  Lastly, we also calculate the mass-normalized number of blue GCs ($T_{\rm blue}$), a quantity particularly useful in understanding the first generation of GC formation in giant galaxies \citep{rh05,br06}.  For each galaxy, we adopt the estimate of the fraction of blue GCs determined in Section~\ref{sec_color} for our calculation of $T_{\rm blue}$.

We discuss the results for the individual galaxies below, comparing our results to previous studies where appropriate.  Our final values of $N_{\rm GC}$, $S_N$, $T$, and $T_{\rm blue}$ are listed in Table~\ref{tbl_ngc}. 

{\it NGC~5866}. -- For NGC~5866, we integrated the best-fit power law profile from $r=0\farcm17$ to $r=11\farcm0$ resulting in 268 GCs.  Inward of $r=0\farcm17$, we estimate an additional 71 GCs would contribute to $N_{\rm GC}$ based on the average of a flat profile and the continuation of the power law to $r = 0$.  Combining these results yields a final estimate of the total number of GCs of $N_{\rm GC} = 340 \pm 80$.  For comparison, CBR07 estimated a total number of GCs of $\sim 400$ from their {\it HST} study, in good agreement with our results. Using the total absolute $V$-band magnitude and galaxy (stellar) mass from Table~\ref{tbl_targets}, we find specific frequencies of $S_N=1.3\pm0.4$ and $T=1.9\pm0.5$.  Using the blue GC fraction of 0.74 determined in Section~\ref{sec_color}, we find a mass-normalized number of blue GCs of $T_{\rm blue} = 1.4 \pm 0.4$.  
Considering the various sources of error $N_{\rm GC}$, we find that the number of GCs assumed for the inner region of the profile is the dominant source of uncertainty in $N_{\rm GC}$ for NGC~5866.  

{\it NGC~5813}. --  For NGC~5813, we integrated the best-fit de Vaucouleurs profile from $r=0\farcm44$ to $r=13\farcm44$ resulting in 2743 GCs.  Assuming the average of a flat radial profile and a continuing de Vaucouleurs profile to $r=0$, we estimate an additional 146 GCs would contribute to $N_{\rm GC}$ from the inner, unmeasurable region of the GC system.  Combining these results gives a total number of GCs $N_{\rm GC}=2900 \pm 400$.  Because the formal error analysis resulted in a fractional uncertainty ${\sigma}_{N_{\rm GC}}/N_{\rm GC} < 10\%$ (due to the relatively low uncertainty in the GCLF coverage), we ultimately adopted an error on $N_{\rm GC}$ commensurate with the typical fractional uncertainty from our wide-field imaging survey ($\sim 10\%-30\%$).  Using the total absolute $V$-band magnitude and galaxy (stellar) mass from Table~\ref{tbl_targets}, we find specific frequencies of $S_N=3.6\pm0.8$ and $T=4.2\pm0.9$.  Using the blue GC fraction of $0.68$ determined in Section~\ref{sec_color}, we find a mass-normalized number of blue GCs of $T_{\rm blue} = 2.8 \pm 0.6$ . \citet{ha81} found $N_{\rm GC} = 2400 \pm 700$, and although their result is consistent with ours, their uncertainties are nearly two times as large as ours.  \citet{ho95} estimated $N_{\rm GC} = 1650 \pm 400$, approximately $3\sigma$ smaller than our estimate.  They find that the radial surface density of their GC candidates extends to only a projected radius of $3\farcm2$ (compared to our estimate of $r\sim13\arcmin$), likely contributing to their lower estimate of $N_{\rm GC}$.

{\it NGC~4762}. -- For NGC~4762, we integrated the best-fit de Vaucouleurs profile from $r=0\farcm1$  to $r=5\farcm5$ resulting in 261 GCs.  Inwards of $r=0\farcm1$, we estimate that an additional 9 GCs would contribute to the total based on the average of a flat inner profile and the continuation of the de Vaucouleurs profile inward to $r=0$.  We therefore find a total number of GCs of $N_{\rm GC} = 270 \pm 30$, where the dominant source of error is the uncertainty in the GCLF correction.  We find specific frequencies of $S_N = 0.9 \pm 0.2$ and $T = 1.4 \pm 0.3$ using our estimate of $N_{\rm GC}$, the galaxy absolute $V$-band magnitude, and the galaxy (stellar) mass from Table~\ref{tbl_targets}.  Using the blue GC fraction of $0.40$ determined in Section~\ref{sec_color}, we find a mass-normalized number of blue GCs of $T_{\rm blue} = 0.6 \pm 0.1$.  For comparison, \citet{pe08} estimate a total number of GCs in NGC~4762 of $211 \pm 34$ from their analysis of the ACSVCS data. Although  \citet{pe08} do include an extrapolation of their best-fit S\'ersic radial surface density profile to account for the limited ACS/WFC field of view, it is unclear to what radius they integrated the profile to derive $N_{\rm GC}$.  Examining our radial profile integration results, we find that the surface density profile outside a projected radius of $\sim 2\farcm7$ (the limit of the ACS/WFC) contributes $\sim 50$ GCs to $N_{\rm GC}$ -- nearly 20\% of the total.  Not surprisingly then, differences and/or uncertainties in the assumed radial extent and surface density profile shapes can contribute significantly to systematic uncertainties in the total number of GCs.

{\it NGC~4754}. -- For NGC~4754, we integrated the best-fit de Vaucouleurs profile from $r=0\farcm09$  to $r=3\farcm5$ resulting in 113 GCs. Inwards of $r=0\farcm09$, we estimate that an additional 2 GCs would contribute to the total based on the average of flat inner profile and the continuation of the de Vaucouleurs profile inward to $r=0$.  We therefore find a total number of GCs of $N_{\rm GC} = 115 \pm 15$, where the dominant source of uncertainty is the Poisson errors on the number of GCs and contaminating objects.  We find specific frequencies of $S_N = 0.6 \pm 0.1$ and $T = 0.9 \pm 0.2$ using our estimate of $N_{\rm GC}$ and the galaxy absolute magnitude and (stellar) mass from Table~\ref{tbl_targets}.  Using the blue GC fraction of $0.40$ determined in Section~\ref{sec_color}, we find a mass-normalized blue GC fraction of $T_{\rm blue} = 0.4 \pm 0.1$.   Based on their ACSVCS study of NGC~4754, \citet{pe08} estimate a total number of GCs in NGC~4754 of $103 \pm 17$ (based on an extrapolated S\'ersic fit), in good agreement with our results.

{\it NGC~3384}. -- For NGC~3384, we integrated the best-fit power law profile from $r=0\farcm59$  to $r=5\farcm6$ resulting in 106 GCs.  Inwards of $r=0\farcm59$, we estimate than an additional 16 GCs would contribute to the total based on the average of the flat inner profile and the continuation of the best-fit power law profile inward to $r=0$.  We thus find a total number of GCs $N_{\rm GC} = 120 \pm 30$. 
We found that all terms in the error analysis contributed nearly equally; no source of uncertainty dominates the error in $N_{\rm GC}$.  We find specific frequencies of $S_N = 0.8 \pm 0.2$ and $T = 1.2 \pm 0.4$ using our estimate of $N_{\rm GC}$ and the galaxy absolute magnitude and (stellar) mass from Table~\ref{tbl_targets}.  Using the blue GC fraction of $0.60$ determined in Section~\ref{sec_color}, we find a mass-normalized blue GC fraction of $T_{\rm blue} = 0.7 \pm 0.2$.  For comparison, \citet{ha81} estimated $N_{\rm GC} = 110 \pm 60$ for NGC~3384 from their analysis of the N3379 field.  Although their photographic study is in good agreement with our result, our three-filter CCD observations and analysis methods have reduced the uncertainty in $N_{\rm GC}$ by a factor of two.

\subsubsection{Comparison of Results to Other Giant Galaxies}\label{sec_survey}

A primary goal of our extragalactic GC system studies is to better quantify how the global properties of GC systems vary with host galaxy properties such as luminosity, environment, and morphology.  Our analysis of the five early-type galaxies in this study adds one elliptical and four lenticular galaxies to our larger wide-field imaging survey.  This doubles the number of lenticular galaxies in our survey sample and results in a comparable numbers of ellipiticals and S0s.  In combination with the spiral galaxies, our current sample of giant galaxies with well-measured total numbers of GCs and specific frequencies consists of 26 galaxies: 11 spirals, 8 lenticulars, and 7 ellipticals.  Twenty-one of these are from 
our wide-field imaging survey and five galaxies are from the literature, including the Milky Way and M31.  The results for 22 of these galaxies are summarized in Y12 and this work adds an additional four galaxies. (The results for NGC~3384 were included in the analysis by Y12.)   Literature references and values for $N_{\rm GC}$, $S_N$, $T$, and $T_{\rm blue}$ for the 22 galaxy sample can be found in Table 10 of Y12.  Although future study will perform a more detailed analysis of the full survey sample (Hargis \& Rhode 2013, in preparation), here we compare values of $S_N$ and $T$ by host galaxy morphological type.

To investigate the variation in specific frequencies with galaxy morphology, we calculated the weighted mean values of $S_N$ and $T$ for the sample of 26 elliptical, spiral, and lenticular galaxies.  The sample of spiral galaxies is identical to that presented in Y12; for 11 galaxies (including the Milky Way and M31) they find weighted mean specific frequencies of $S_N = 0.6 \pm 0.1$ and $T=1.2 \pm 0.1$.  For the sample of seven elliptical galaxies we find weighted means of $S_N = 1.8 \pm 0.1$ and $T=2.1 \pm 0.2$, consistent with the combined analysis of 11 E/S0 galaxies by Y12.  For the sample of eight lenticular galaxies we find weighted means of $S_N =0.9 \pm 0.1$ and $T= 1.5 \pm 0.1$.  Consistent with previous results on giant galaxy GC systems, we find that ellipticals have larger luminosity- and mass-normalized specific frequencies than spirals.  Comparing the lenticulars to spirals and ellipticals, we find that on average, S0s have GC specific frequencies more comparable to spirals than to ellipticals.  The difference in mean values of $S_N$ and $T$ between spirals and lenticulars ($\sim 0.3$) is only marginally larger than the formal uncertainties in the weighted means; the mean values are consistent within $3\sigma$.  For the galaxies in this work, we find that the four lenticulars occupy a range of specific frequencies consistent with both the weighted means of spirals and S0s. Quantifying the degree to which the GC systems of S0s and spirals are statistically different (or similar) will require a larger sample of galaxies with well-measured global properties.  

\section{Summary and Conclusions}\label{sec_conclusions}

In this study we investigated the global properties of the GC systems of five moderate-luminosity ($M_V ~\sim -21$ to $-22$), early-type giant galaxies from our wide-field imaging survey program.  We used KPNO 4-m MOSAIC $BVR$ imaging to trace the GC systems to large projected radii. 
The final results of our study are given in Table~\ref{tbl_ngc}. Here we summarize our main results:

\begin{enumerate}

\item We constructed radial surface density profiles for the GC systems for the target galaxies using the MOSAIC imaging data and, where available, complementary archival and published {\it HST} data.  While the large spatial coverage of the MOSAIC imager allows us to trace the GC population to galactocentric radii of $\sim 150-200$ kpc (for the distances of our target galaxies), the higher spatial resolution of {\it HST} instruments allows us to probe the inner several hundred parsecs.  

\item We integrate the best-fit radial surface density profiles to derive estimates of the total number of GCs ($N_{\rm GC}$).  We find total numbers of GCs ranging from $N_{\rm GC}\sim 120$ for the two lowest luminosity galaxies to $N_{\rm GC}\sim 2900$ for the most luminous galaxy in our study.  We find $V$-band normalized specific frequencies $S_N$ ranging from $0.6$ to $3.6$ and stellar mass-normalized specific frequencies of $T$ ranging from $0.9$ to $4.2$.  Analyzing a larger sample of 26 galaxies with well-measured total numbers and specific frequencies, we find that spiral and lenticular galaxies have statistically consistent (weighted) mean values of $S_N$ and $T$.  The four S0 galaxies in this study have values of $S_N$ and $T$ comparable to the mean values of both spiral and lenticular galaxies.  

\item  We studied the color distributions and color gradients of the
galaxies' GC systems using the MOSAIC imaging data.  Only NGC~5813
shows strong evidence ($99.99\%$ confidence
level) of color bimodality.  NGC~5866 shows some evidence for color
bimodality, although at a lower level of statistical significance
($97\%$ confidence level).  NGC~5813 has sufficient numbers of GC
candidates to investigate the spatial distributions of the red
(metal-rich) and blue (metal-poor) GC subpopulations.  The red GC
subpopulation is more centrally concentrated about the galaxy center
than the blue GC subpopulation and a statistically significant $B-R$
color gradient is found in the inner $6\arcmin$ ($\sim 55$ kpc) of the GC system.  No
statistically significant color gradients are found in the GC systems
of the other target galaxies.

\end{enumerate}

The primary goal of this work has been to measure the global properties of the GC systems for a number of intermediate-luminosity, early-type giant galaxies.  While a fair number of the most massive early-type galaxies have been studied using wide-field imaging techniques \citep{di03,ha04,rh01,rh04,ta06,fo11}, studies with large spatial coverage are rare for moderate-luminosity galaxies.   With the addition of these five galaxies to our larger survey sample, a future study (Hargis \& Rhode 2013, in preparation) will present the results of a multi-variate analysis of the current survey sample of $\sim 30$ galaxies.

\acknowledgments

This research was supported by an NSF Faculty Early Career Development(CAREER) award (AST-0847109) to K.L.R.  We are grateful to the staff
of the WIYN Observatory and Kitt Peak National Observatory for their
assistance during the observing runs.  J.R.H. would like to thank NOAO for financial support through the thesis observing program. J.R.H. would like to thank
Steven Janowiecki and Liese van Zee for useful conversations during
the course of this work. We would like to thank the anonymous referee
for his/her useful comments which have improved the quality of the paper. 
This research has made use of the NASA/IPAC Extragalactic
Database (NED) which is operated by the Jet Propulsion Laboratory,
California Institute of Technology, under contract with the National
Aeronautics and Space Administration.



\clearpage


\clearpage

\clearpage

\clearpage

\begin{deluxetable}{llcccccccc}
\rotate
\tablecolumns{10}
\tablewidth{0pc}
\tabletypesize{\scriptsize}
\tablecaption{Basic Properties of Target Early-Type Galaxies\label{tbl_targets}}
\tablehead{
\colhead{Name (VCC ID)} & \colhead{Type} & \colhead{$M_V^T$} & \colhead{$m-M$} & \colhead{Distance (Mpc)} & \colhead{$V_T^0$} & \colhead{$A_V$} & \colhead{$v_{\rm helio}$ (${\rm km~s}^{-1}$)} & \colhead{$\log{M/M_\odot}$} & \colhead{Environment}
}
\startdata
NGC~5866            &  SA0  &   $-21.1$ & $30.93 \pm 0.12$   & $15.3$   &  $9.85$\tablenotemark{1}  & $0.044$  & $672$  &  $11.2$ & NGC~5866 Group \\
NGC~5813            &  E1   &   $-22.3$ & $32.54 \pm 0.18$   & $32.2$   &  $10.27$\tablenotemark{1} & $0.189$  & $1972$ &  $11.8$ & NGC~5846 Group \\
NGC~4762 (VCC~2095) &  SB0  &   $-21.2$ & $31.35 \pm 0.15$\tablenotemark{2}   & $18.6$   &  $10.16$ & $0.086$  & $984$  &  $11.3$ & Virgo Cluster  \\
NGC~4754 (VCC~2092) &  SB0  &   $-20.7$ & $31.13 \pm 0.14$   & $16.8$   &  $10.43$ & $0.086$  & $1347$ &  $11.1$ & Virgo Cluster  \\
NGC~3384            &  SB0  &   $-20.5$ & $30.32 \pm 0.14$   & $11.6$   &  $9.84$  & $0.081$  & $704$  &  $11.0$ & Leo I Group    
\enddata
\tablecomments{Galaxy morphological types are from \citet{RC3}.  $M_V^T$ are computed by combining $m-M$ with $V_T^0$.  $V_T^0$ are from \citet{RC3} except where noted. $m-M$ are from \citet{to01}, except where noted.  Galactic extinction values $A_V$ are from \citet{sfd98}. Heliocentric radial velocities $v_{\rm helio}$ are from the NASA/IPAC Extragalactic Database (NED). Galaxy stellar masses $\log{M/M_\odot}$ are computed from $M_V^T$ and assumed $(M/L)_V$ from \citet{ze93} of $7.6$ for lenticulars and $10.0$ for ellipticals.
}
\tablenotetext{1}{Computed from $V_T$ from \citet{RC3} and the adopted $A_V$.}
\tablenotetext{2}{Computed by combining the 3K CMB recession velocity of $1302 \pm 25~ {\rm km ~s}^{-1}$ (NED) with $H_0 = 70 \pm 5~{\rm km~ s}^{-1}~{\rm Mpc}^{-1}$}
\end{deluxetable}

\begin{deluxetable}{lcccc}
\tablecolumns{5}
\tablewidth{0pc}
\tablecaption{Mosaic Observations of Target Galaxies.\label{tbl_observations}}
\tablehead{
\colhead{Name} & \colhead{Date} & \multicolumn{3}{c}{Exposure Times (s)}\\
\cline{3-5} \\
\colhead{} & \colhead{} & \colhead{$B$} & \colhead{$V$} & \colhead{$R$}
}
\startdata
NGC~5866 & May 2010 & $4\times800;1\times600$ & $5\times700$ & $3\times600$\\
NGC~5813 & May 2010 & $4\times1260$ & $5\times1100$ & $4\times1080$ \\
NGC~4762/4754 & May 2010 & $5\times1260$ & $4\times1200$ & $3\times1080$ \\
NGC~4216 & May 2010 & $5\times1260$ & $5\times1200$ & $5\times1080$ \\
NGC~3384 & March 1999 & $5\times660$ & $5\times540$ & $5\times480$ 
\enddata
\end{deluxetable}

\begin{deluxetable}{lcccccccccccc} 
\tablecolumns{13} 
\tablewidth{0pc} 
\tablecaption{50\% and 90\% Completeness Limits for Mosaic Data \label{tbl_completeness}} 
\tablehead{ 
\colhead{Image} & \colhead{} & \multicolumn{2}{c}{NGC~5866} &   \colhead{}   & 
                               \multicolumn{2}{c}{NGC~5813} &   \colhead{}   &
                               \multicolumn{2}{c}{NGC~4762/4754} &   \colhead{}   & 
                               \multicolumn{2}{c}{NGC~3384} \\
\cline{3-4} \cline{6-7} \cline{9-10} \cline{12-13}  \\
\colhead{} & \colhead{} &      \colhead{50\%}   & \colhead{90\%}
           & \colhead{} &      \colhead{50\%}   & \colhead{90\%}   
           & \colhead{} &      \colhead{50\%}   & \colhead{90\%} 
           & \colhead{} &      \colhead{50\%}   & \colhead{90\%}
}
\startdata
$B$ & &  23.80 &  23.46 & &  25.13 & 24.80 & &     24.63   & 24.13     & & 23.56 & 22.60  \\
$V$ & &  24.03 &  23.63 & &  24.23 & 23.85 & &     23.97   & 23.62     & & 24.09 & 23.19  \\
$R$ & &  23.46 &  23.03 & &  24.07 & 23.48 & &     23.17   & 22.82     & & 23.40 & 22.84  \\ 
\enddata
\end{deluxetable}

\begin{deluxetable}{lcccc}
\tablecolumns{5}
\tablewidth{0pc}
\tablecaption{{\it HST} Observations Analyzed for this Study \label{tbl_hst}}
\tablehead{
\colhead{Proposal ID} &  \colhead{PI} & \colhead{Instrument} &  \colhead{Exposure Time (s)} & \colhead{Filter}\\
\cline{1-5}
\multicolumn{5}{c}{NGC~5866} 
}
\startdata
$10705$ & Noll & ACS/WFC & 3900 & F435W \\
$10705$ & Noll & ACS/WFC & 2800 & F555W \\
$10705$ & Noll & ACS/WFC & 2200 & F625W \\
$978$   & Ho   & ACS/WFC & 120  & F814W \\
\cutinhead{NGC~5813}
$5454$  & Franx& WFPC2   & 1000 & F555W \\
$5454$  & Franx& WFPC2   & 460  & F814W \\
$6357$  & Jaffe& WFPC2   & 1000 & F702W \\
\cutinhead{NGC~4762} 
$9401$  & C\^ote& ACS/WFC& 750  & F475W \\
$9401$  & C\^ote& ACS/WFC& 1210 & F850LP \\
\cutinhead{NGC~4754}
$9401$  & C\^ote& ACS/WFC& 750  & F475W \\
$9401$  & C\^ote& ACS/WFC& 1210 & F850LP \\
\enddata
\tablecomments{{\it HST} observations of NGC~3384 were analyzed in RZ04.  See their Table~5 for details.}
\end{deluxetable}


\begin{deluxetable}{lcccc}
\tablecolumns{5}
\tablewidth{0pc}
\tablecaption{Corrected Radial Surface Density Profile for GC Candidates in NGC~5866 \label{tbl_n5866_prof}}
\tablehead{
\colhead{Projected Radius (arcmin)}  &   \colhead{Surface Density (${\rm arcmin}^{-2}$)} & \colhead{Fractional Coverage} & \colhead{Source}
}
\startdata
 0.4 & $32.46 \pm 7.26$ &  0.89 & $HST$ \\  
 0.8 & $16.55 \pm 3.38$ &  1.00 & $HST$ \\  
 1.2 & $ 6.81 \pm 1.76$ &  0.99 & $HST$ \\  
 1.6 & $ 3.90 \pm 1.30$ &  0.77 & $HST$ \\  
 2.0 & $ 1.88 \pm 1.33$ &  0.28 & $HST$ \\  
 2.6 & $ 1.42 \pm 0.41$ &  0.82 & Mosaic \\  
 4.9 & $ 0.66 \pm 0.20$ &  1.00 & Mosaic \\  
 7.3 & $ 0.23 \pm 0.14$ &  0.94 & Mosaic \\  
 9.8 & $ 0.13 \pm 0.12$ &  0.78 & Mosaic \\  
12.3 & $ 0.08 \pm 0.11$ &  0.72 & Mosaic \\  
14.8 & $ 0.06 \pm 0.11$ &  0.65 & Mosaic \\  
17.2 & $ 0.09 \pm 0.11$ &  0.50 & Mosaic \\  
19.7 & $-0.10 \pm 0.11$ &  0.33 & Mosaic \\  
\enddata
\tablecomments{Negative surface densities can occur due to the
 application of the contamination correction.}
\end{deluxetable}

\begin{deluxetable}{lcccc}
\tablecolumns{5}
\tablewidth{0pc}
\tablecaption{Corrected Radial Surface Density Profile for GC Candidates in NGC~5813 \label{tbl_n5813_prof}}
\tablehead{
\colhead{Projected Radius (arcmin)}  &   \colhead{Surface Density (${\rm arcmin}^{-2}$)} & \colhead{Fractional Coverage} & \colhead{Source}
}
\startdata
 1.1 & $77.27 \pm 8.82$ &  0.90 & Mosaic \\  
 2.0 & $35.88 \pm 4.18$ &  0.94 & Mosaic \\  
 3.0 & $18.51 \pm 2.66$ &  0.86 & Mosaic \\  
 4.0 & $10.24 \pm 1.73$ &  0.96 & Mosaic \\  
 5.0 & $ 6.93 \pm 1.35$ &  0.96 & Mosaic \\  
 6.0 & $ 5.38 \pm 1.14$ &  0.95 & Mosaic \\  
 7.0 & $ 3.78 \pm 0.96$ &  0.96 & Mosaic \\  
 8.0 & $ 2.44 \pm 0.82$ &  0.96 & Mosaic \\  
 8.9 & $ 1.72 \pm 0.74$ &  0.92 & Mosaic \\  
10.0 & $ 0.69 \pm 0.63$ &  0.95 & Mosaic \\  
10.9 & $ 1.36 \pm 0.64$ &  0.97 & Mosaic \\  
11.9 & $ 0.81 \pm 0.59$ &  0.94 & Mosaic \\  
12.9 & $ 0.27 \pm 0.53$ &  0.96 & Mosaic \\  
13.9 & $ 1.02 \pm 0.58$ &  0.87 & Mosaic \\  
14.9 & $-0.09 \pm 0.52$ &  0.79 & Mosaic \\  
15.9 & $ 0.47 \pm 0.54$ &  0.78 & Mosaic \\  
16.9 & $ 0.22 \pm 0.52$ &  0.76 & Mosaic \\  
17.9 & $ 0.51 \pm 0.56$ &  0.64 & Mosaic \\  
18.9 & $-0.86 \pm 0.51$ &  0.50 & Mosaic \\  
19.9 & $ 0.27 \pm 0.64$ &  0.41 & Mosaic \\  
20.9 & $-0.18 \pm 0.65$ &  0.34 & Mosaic \\  
21.9 & $-0.44 \pm 0.65$ &  0.29 & Mosaic \\  
22.9 & $ 0.08 \pm 0.82$ &  0.20 & Mosaic \\  
23.9 & $ 0.35 \pm 0.98$ &  0.14 & Mosaic \\  
24.9 & $ 0.32 \pm 1.12$ &  0.10 & Mosaic \\  
25.9 & $-0.69 \pm 1.17$ &  0.07 & Mosaic \\  
\enddata
\tablecomments{Negative surface densities can occur due to the
 application of the contamination correction.}
\end{deluxetable}

\begin{deluxetable}{lcccc}
\tablecolumns{5}
\tablewidth{0pc}
\tablecaption{Corrected Radial Surface Density Profile for GC Candidates in NGC~4762 \label{tbl_n4762_prof}}
\tablehead{
\colhead{Projected Radius (arcmin)}  &   \colhead{Surface Density (${\rm arcmin}^{-2}$)} & \colhead{Fractional Coverage} & \colhead{Source}
}
\startdata
 0.5 & $42.62 \pm 7.03$ &  0.97 & $HST$ \\  
 1.0 & $15.44 \pm 2.69$ &  0.99 & $HST$ \\  
 1.3 & $ 9.08 \pm 2.21$ &  0.77 & Mosaic \\  
 1.6 & $ 5.68 \pm 1.55$ &  0.74 & $HST$ \\  
 2.1 & $ 3.05 \pm 1.88$ &  0.21 & $HST$ \\  
 2.4 & $ 1.55 \pm 0.79$ &  0.79 & Mosaic \\  
 3.7 & $ 0.78 \pm 0.52$ &  0.88 & Mosaic \\  
 4.9 & $ 0.15 \pm 0.37$ &  0.94 & Mosaic \\  
 6.2 & $ 0.17 \pm 0.33$ &  0.97 & Mosaic \\  
 7.4 & $-0.12 \pm 0.28$ &  0.92 & Mosaic \\  
 8.6 & $ 0.62 \pm 0.33$ &  0.88 & Mosaic \\  
 9.9 & $ 0.15 \pm 0.28$ &  0.87 & Mosaic \\  
\enddata
\tablecomments{Negative surface densities can occur due to the
 application of the contamination correction.}
\end{deluxetable}

\begin{deluxetable}{lcccc}
\tablecolumns{5}
\tablewidth{0pc}
\tablecaption{Corrected Radial Surface Density Profile for GC Candidates in NGC~4754 \label{tbl_n4754_prof}}
\tablehead{
\colhead{Projected Radius (arcmin)}  &   \colhead{Surface Density (${\rm arcmin}^{-2}$)} & \colhead{Fractional Coverage} & \colhead{Source}
}
\startdata
 0.2 & $53.00 \pm 18.93$ & 1.00 & $HST$ \\  
 0.5 & $16.22 \pm 3.85$ &  0.97 & $HST$ \\  
 0.9 & $ 5.20 \pm 2.39$ &  0.81 & Mosaic \\  
 1.4 & $ 5.05 \pm 0.86$ &  0.82 & $HST$ \\  
 1.6 & $ 1.70 \pm 1.07$ &  0.95 & Mosaic \\  
 2.3 & $ 0.79 \pm 0.73$ &  0.94 & Mosaic \\  
 3.1 & $ 0.33 \pm 0.56$ &  0.94 & Mosaic \\  
 3.8 & $ 0.02 \pm 0.45$ &  0.96 & Mosaic \\  
 4.6 & $ 0.49 \pm 0.48$ &  0.97 & Mosaic \\  
 5.3 & $-0.01 \pm 0.39$ &  0.91 & Mosaic \\  
 6.1 & $ 0.28 \pm 0.42$ &  0.85 & Mosaic \\  
 6.8 & $ 0.36 \pm 0.41$ &  0.83 & Mosaic \\  
 7.6 & $ 0.85 \pm 0.45$ &  0.81 & Mosaic \\  
 8.3 & $ 0.29 \pm 0.38$ &  0.76 & Mosaic \\  
\enddata
\tablecomments{Negative surface densities can occur due to the
 application of the contamination correction.}
\end{deluxetable}

\begin{deluxetable}{lcccc}
\tablecolumns{5}
\tablewidth{0pc}
\tablecaption{Corrected Radial Surface Density Profile for GC Candidates in NGC~3384 \label{tbl_n3384_prof}}
\tablehead{
\colhead{Projected Radius (arcmin)}  &   \colhead{Surface Density (${\rm arcmin}^{-2}$)} & \colhead{Fractional Coverage} & \colhead{Source}
}
\startdata
 1.2 & $ 3.48 \pm 1.24$ &  0.85 & Mosaic \\  
 2.1 & $ 1.70 \pm 0.61$ &  0.98 & Mosaic \\  
 3.1 & $ 1.36 \pm 0.52$ &  0.79 & Mosaic \\  
 4.1 & $ 0.34 \pm 0.31$ &  0.70 & Mosaic \\  
 5.1 & $ 0.00 \pm 0.22$ &  0.68 & Mosaic \\  
 6.1 & $ 0.03 \pm 0.21$ &  0.66 & Mosaic \\  
 7.1 & $ 0.11 \pm 0.20$ &  0.68 & Mosaic \\  
 8.1 & $ 0.23 \pm 0.21$ &  0.69 & Mosaic \\  
 9.1 & $ 0.03 \pm 0.17$ &  0.71 & Mosaic \\  
10.1 & $ 0.17 \pm 0.18$ &  0.72 & Mosaic \\  
11.1 & $-0.02 \pm 0.14$ &  0.73 & Mosaic \\  
12.1 & $ 0.17 \pm 0.16$ &  0.77 & Mosaic \\  
13.1 & $-0.02 \pm 0.13$ &  0.76 & Mosaic \\  
14.1 & $-0.08 \pm 0.12$ &  0.75 & Mosaic \\  
15.1 & $ 0.25 \pm 0.16$ &  0.72 & Mosaic \\  
16.1 & $-0.13 \pm 0.11$ &  0.61 & Mosaic \\  
17.1 & $ 0.11 \pm 0.15$ &  0.56 & Mosaic \\  
18.1 & $-0.11 \pm 0.12$ &  0.51 & Mosaic \\  
\enddata
\tablecomments{Negative surface densities can occur due to the
 application of the contamination correction.}
\end{deluxetable}


\begin{deluxetable}{lcccccccc}
\tablecolumns{9}
\tablewidth{0pc}
\tablecaption{GC Radial Surface Density Profile Fit Parameters \label{tbl_prof_fits}}
\tablehead{
\colhead{} &  \colhead{} & \multicolumn{3}{c}{Power Law}  & \colhead{} & \multicolumn{3}{c}{de Vaucouleurs Law} \\
\cline{3-5} \cline{7-9}\\
\colhead{Galaxy} & \colhead{} & \colhead{$a_0$} & \colhead{$a_1$} & \colhead{${\chi}^2/\nu$} & \colhead{} & \colhead{$a_0$} & \colhead{$a_1$} & \colhead{${\chi}^2/\nu$}
}
\startdata
NGC~5866  & & $0.96 \pm 0.04$ & $-1.76 \pm 0.09$ & $0.47$ &  & $3.52 \pm 0.16$ & $-2.52 \pm 0.14$ & $0.70$ \\
NGC~5813  & & $2.02 \pm 0.04$ & $-1.74 \pm 0.07$ & $0.86$ &  & $4.18 \pm 0.12$ & $-2.23 \pm 0.09$ & $0.33$ \\
NGC~4762  & & $1.10 \pm 0.04$ & $-1.86 \pm 0.14$ & $1.02$ &  & $4.23 \pm 0.23$ & $-3.09 \pm 0.23$ & $0.52$ \\
NGC~4754  & & $0.78 \pm 0.05$ & $-1.41 \pm 0.13$ & $1.36$ &  & $3.44 \pm 0.24$ & $-2.61 \pm 0.24$ & $0.97$ \\
NGC~3384  & & $0.65 \pm 0.11$ & $-1.29 \pm 0.17$ & $0.88$ &  & $2.06 \pm 0.28$ & $-1.51 \pm 0.20$ & $1.02$ 
\enddata
\end{deluxetable}

\begin{deluxetable}{lcccccccc}
\rotate
\tablecolumns{9}
\tablewidth{0pc}
\tablecaption{Total Numbers of GCs and Specific Frequencies of GC Systems for Target Galaxies \label{tbl_ngc}}
\tablehead{
\colhead{Galaxy} & \colhead{$N_{\rm GC}$} & \colhead{$S_N$} & \colhead{$T$} & \colhead{$T_{\rm blue}$} & \colhead{$f_{\rm blue}$} & \colhead{Color Bimodality} & \colhead{Color Gradient} & \colhead{Radial Extent} \\
\colhead{}       & \colhead{}             & \colhead{}      & \colhead{}    & \colhead{}               & \colhead{}               & \colhead{(Yes/No)} & \colhead{(Yes/No)} & \colhead{(kpc)} 
}
\startdata
NGC~5866 & $340 \pm 80$ & $1.3 \pm 0.4$ & $1.9 \pm 0.5$ & $1.4 \pm 0.4$ & $0.74$ & Y  & N  & $44 \pm 11$  \\
NGC~5813 & $2900\pm400$ & $3.6 \pm 0.8$ & $4.2 \pm 0.9$ & $2.8 \pm 0.6$ & $0.68$ & Y  & Y  & $120 \pm 14$ \\
NGC~4762 & $270 \pm 30$ & $0.9 \pm 0.2$ & $1.4 \pm 0.3$ & $0.6 \pm 0.1$ & $0.40$ & N  & N  & $27 \pm 7$  \\
NGC~4754 & $115 \pm 15$ & $0.6 \pm 0.1$ & $0.9 \pm 0.2$ & $0.4 \pm 0.1$ & $0.40$ & N  & N  & $15 \pm 4$  \\
NGC~3384 & $120 \pm 30$ & $0.8 \pm 0.2$ & $1.2 \pm 0.4$ & $0.7 \pm 0.2$ & $0.60$ & N  & N  & $17 \pm 4$
\enddata
\end{deluxetable}

\end{document}